\documentclass{andp2012}
\usepackage[english]{babel}
\usepackage{bm}
\usepackage{color}
\usepackage{widetext}
\usepackage{caption}
\usepackage{subcaption}
\usepackage{graphicx}
\DeclareMathAlphabet\mathbfcal{OMS}{cmsy}{b}{n}

\def\ketm#1{  \left\vert  #1   \right\rangle   }

\def\sprm#1#2{  \left\langle #1 \left\vert \right. #2 \right\rangle   }

\def\mem#1#2#3{  \left\langle #1 \left\vert  #2 \right\vert #3 \right\rangle   }

\def\rmem#1#2#3{  \left\langle #1 \left\vert \left\vert  #2
                  \right\vert \right\vert #3 \right\rangle   }

\def\sixjm#1#2#3#4#5#6{  \left\{ \begin{array}{ccc}
                                               #1 & #2 & #3  \\
                                               #4 & #5 & #6
                     \end{array} \right\}   }

\def\asc#1{{\color{black} #1}}


\title{Parity--violation studies with partially stripped ions}

\author[J. Richter]{Jan Richter\inst{1, 2}}
\author[A.\,V. Maiorova]{Anna V. Maiorova\inst{3,4}}
\author[A.\,V. Viatkina]{Anna V. Viatkina\inst{1, 2, 5, 6}}
\author[D. Budker]{Dmitry Budker\inst{5, 6, 7}}
\author[A. Surzhykov]{Andrey Surzhykov\inst{1, 2, 8, }\footnote{Corresponding author\quad E-mail:~\textsf{andrey.surzhykov@ptb.de}}}

\address[1]{Physikalisch--Technische Bundesanstalt, D--38116 Braunschweig, Germany}
\address[2]{Technische Universit\"at Braunschweig, D--38106 Braunschweig, Germany}
\address[3]{Center for Advanced Studies, Peter the Great St. Petersburg Polytechnic University, 195251 St. Petersburg, Russia}
\address[4]{Petersburg Nuclear Physics Institute of NRC “Kurchatov Institute”, 188300 Gatchina, Russia}
\address[5]{Helmholtz Institute Mainz, GSI Helmholtzzentrum für Schwerionenforschung, 55099 Mainz, Germany}
\address[6]{Johannes Gutenberg University Mainz, 55099 Mainz, Germany}
\address[7]{Department of Physics, University of California, Berkeley, California 94720, USA}
\address[8]{Laboratory for Emerging Nanometrology Braunschweig, D-38106 Braunschweig, Germany}

\shortauthors{J. Richter et al.}

\begin{abstract}
We present a theoretical study of photoexcitation of highly charged ions from their ground states, a process which can be realized at the Gamma Factory at CERN. Special attention is paid to the question of how the excitation rates are affected by the \textit{mixing} of opposite--parity ionic levels, which is induced both by an external electric field and the weak interaction between electrons and the nucleus. In order to reinvestigate this ``Stark--\textit{plus}--weak--interaction'' mixing, well--known in neutral atomic systems, we employ relativistic Dirac's theory. Based on the developed approach, detailed calculations are performed for the 1s$_{1/2} \to$~2s$_{1/2}$ and ${\rm 1s}^2 \; {\rm 2s}_{1/2} \to {\rm 1s}^2 \; {\rm 3s}_{1/2}$ (M1 + parity--violating--E1) transitions in hydrogen-- and lithium--like ions, respectively. In particular, we focus on the difference between the excitation rates obtained for the right-- and left--circularly polarized incident light. This difference arises due to the parity violating mixing of ionic levels and is usually characterized in terms of the \textit{circular--dichroism} parameter. 
We argue that future measurements of circular dichroism, performed with highly charged ions in the SPS or LHC rings, may provide valuable information on the electron--nucleus weak--interaction coupling. 
\end{abstract}
\shortabstract

\begin{document}
\maketitle

%
%
\section{Introduction}

Atomic parity-violation (APV) phenomena arising due to the weak interaction of atomic electrons with nuclei have been the focus of experimental and theoretical research for several decades \cite{Khr91,BoB97,GuL05,BuK08, SaB18}. These studies help to probe the electroweak segment of the Standard Model in the low--energy regime, which makes them complementary to the high--energy particle--physics experiments. Along with neutral atoms and molecules \cite{WoB97,BeW99,DeC08}, \textit{highly charged ions} are considered to be promising candidates for exploring the electron--nucleus weak coupling. The advantages of ions over neutral atomic systems are their much simpler electronic structure and a large overlap of electronic density with the nucleus, which leads to a significant enhancement of APV effects \cite{ScS89,LaN01}. Moreover, by varying nuclear charge $Z$ of an ion and a number of bound electrons, one can ``tune'' energy spectra in such a way that two ionic levels of opposite parity become nearly degenerate. Such a degeneracy further enhances the \textit{mixing} between the opposite--parity levels, caused by the weak interaction. This level mixing is the most pronounced APV effect, and its observation provides unique information about the neutral--current electron--quark coupling constants.    

A number of scenarios have been proposed to observe the PV level mixing in highly charged ions in future storage-ring experiments \cite{ScS89,LaN01,ShV10,BoP11,FeS11}. In most of these scenarios, one would need to observe laser--induced transitions from initially \textit{excited} ionic states. A short lifetime of these excited states, which usually does not exceed 10$^{-12}$\,s for high--$Z$ domain, imposes serious constraints on the practical realization of the proposed schemes. An alternative and promising route for APV studies in storage rings is the photoexcitation of ions from their \textit{ground} states. In the past, however, this method was hindered by the necessity of application of lasers in the x-ray regime to induce ionic transitions. This difficulty will be naturally overcome at the Gamma Factory facility, whose realization is currently under discussion within the framework of the ``Physics Beyond Colliders'' initiative at CERN \cite{JaL18,Kra15,BuC20}. In the Gamma Factory, the ions, moving with a high relativistic Lorentz factor $\gamma = 1/\sqrt{1 - v^2/c^2} \gg 1$, will undergo resonant excitation induced by incident laser radiation. Owing to the Lorentz transformation, the energy of laser photons counter-propagating to the ions will be boosted by a factor of 2$\gamma$ in the ion rest frame, thus enabling x-ray spectroscopy of ionic transitions and, hence, measurement of parity-violation effects \cite{ZoB97}.

In the present theoretical study, we revisit laser excitation of highly charged ions from their ground states, as will be possible at the Gamma Factory, and discuss how this excitation is affected by the parity--violating mixing between ionic levels. Moreover, we show how the system can be \textit{controlled} by applying external electric field. In order to investigate the well--established (for neutral atoms) Stark--PV interference scheme, we apply in Section~\ref{sec:theory} the first--order perturbation theory and obtain the coefficients that describe weak--interaction-- as well as Stark mixing between ionic levels of opposite parity. Moreover, the evaluation of the photoexcitation matrix elements is also briefly reviewed in Section~\ref{subsection:transition_matrix_element}. While the derived expressions are general and can be applied to any ion (or atom), independent on its shell structure, here we apply them to explore the  1s$_{1/2} \to$~2s$_{1/2}$ and ${\rm 1s}^2 \; {\rm 2s_{1/2}} \to {\rm 1s}^2 \; {\rm 3s_{1/2}}$ transitions in hydrogen-- and lithium--like ions, respectively. For both transition types, that are of particular interest for the Gamma--Factory project, we focus on \textit{circular dichroism} that characterizes the relative difference between the probabilities to excite atoms with right-- and left--circularly polarized light. In the absence of external fields, the absolute values of the dichroism do not exceed $10^{-6}$ and change only slightly with the nuclear charge. In contrast, our calculations demonstrate that controllable enhancement of the circular dichroism parameter by a factor of $2$ in hydrogen--like ions and $10^3$ in lithium--like ions can be achieved if ions are exposed to a combination of electric and magnetic fields. It is important to note that for the design of the future experiments, in addition to the dichroism parameter, it will be important to consider the absolute rates of transitions in order to optimize the overall sensitivity \cite{ZoB97}. The summary of these results and a brief outlook are given finally in Section \ref{sec:summary}.

For the representation of the results in Sec.\,\ref{sec:results} we employ the atomic unit of electric field strength given by $\mathcal{E}$(a.u.)\,=\,5.1422\,$\times 10^{11}$\,V/m.

\section{Geometry and parameters of the study}
\label{section:setup}

\begin{figure}
    \centering
    \includegraphics[width=0.98\linewidth]{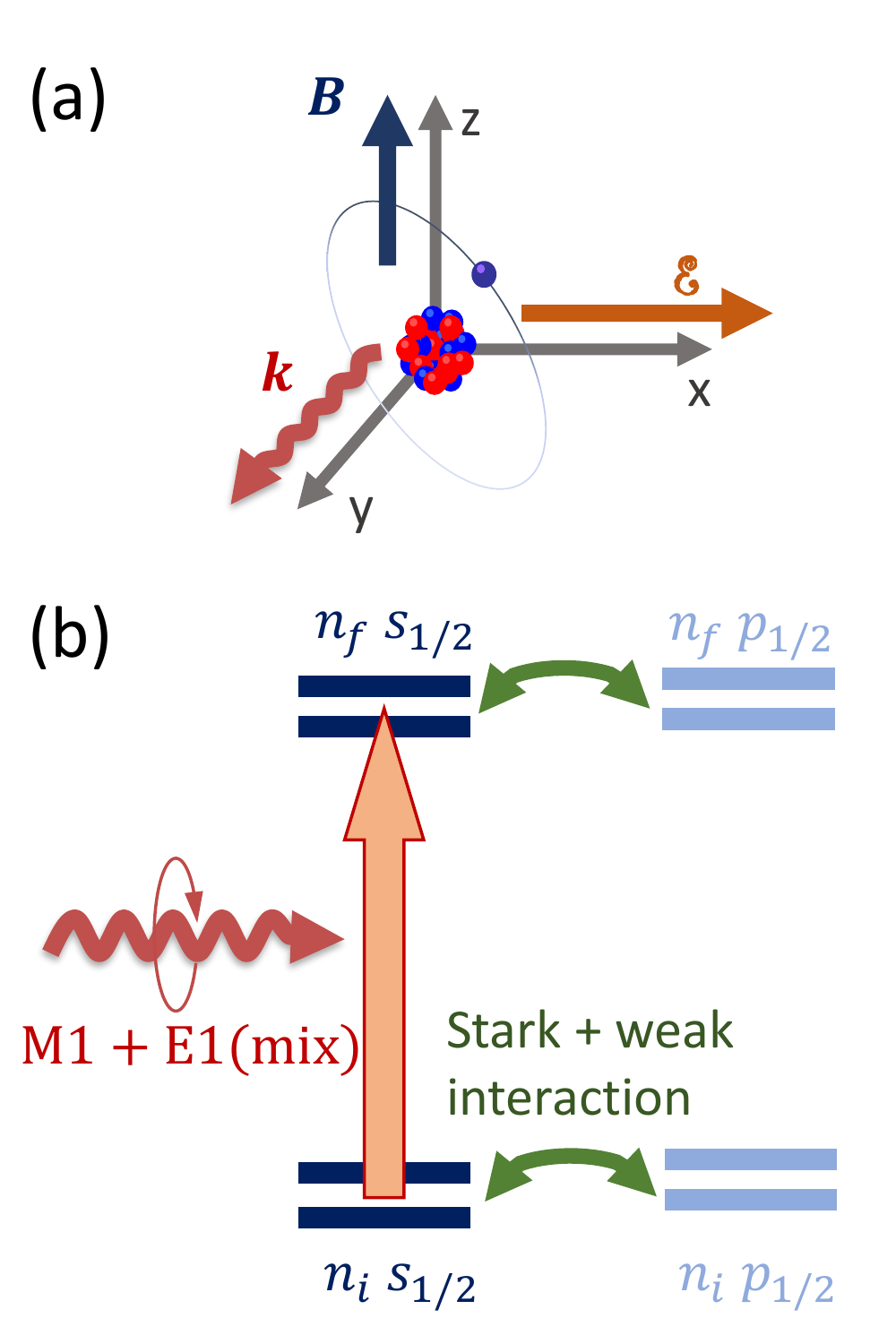}
    \caption{(a) Geometry of the APV study with highly charged ions. Within its rest frame, the ion is exposed to static electric and magnetic fields, whose directions are chosen along the $x$-- and $z$--axis, respectively. Incident circularly polarized light propagates along the $y$--axis. (b) Thanks to the weak--interaction and Stark mixing between $n$s and $n$p ionic states, the $n$s$\to$~$n'$s laser--induced transition may proceed via the leading magnetic dipole (M1) as well as the parity--violating electric dipole (E1) channels. The interference between these two channels results in the sensitivity of the excitation rate to the direction of the circular polarization of incident light.}
    \label{Fig:geometry}
\vspace{-0.5cm}
\end{figure}

Before we present the basic theory to describe the photoexcitation of highly charged ions at the Gamma Factory, let us briefly discuss the setup and geometry of this study. In the present work, we will focus on the 1s$_{1/2} \to$~2s$_{1/2}$ and ${\rm 1s}^2 \; {\rm 2s}_{1/2} \to {\rm 1s}^2 \; {\rm 3s}_{1/2}$ laser--induced transitions in hydrogen-- and lithium--like ions, respectively. Owing to the fact that the weak interaction between electrons and a nucleus as well as the external electric field lead to the \textit{mixing} of $n$s and $n$p ionic states with $n = 2, 3$, these transitions can proceed not only via the leading magnetic dipole (M1) but also via the parity--violating electric dipole (E1) channel, see lower panel of Fig.\,\ref{Fig:geometry}. The interference between the M1 and PV--E1 channels can affect the excitation probabilities. For example, the probabilities to induce $n$s$ \to$~$n'$s transition by right-- and left--circularly polarized light differ from each other. This so--called \textit{circular dichroism}, whose measurements allow one to determine the parameters of the weak interaction \cite{BoB97,BuK08, SaB18}, are the focus of the present study. Below, we investigate how the dichroism can be controlled by applying external electric field and by inducing transitions between magnetic sublevels, $\ketm{{n\rm  s}_{1/2} \, \, \mu_i} \to \ketm{{n'\rm  s}_{1/2} \, \, \mu_f}$. This scenario would require an application of an additional magnetic field in order to introduce Zeeman splitting between sublevels. 

As seen from the discussion above, the study of parity--violating phenomena in the photoexcitation of highly charged ions generally requires the use of external electric and magnetic fields. One has to agree, therefore, about the \textit{geometry} and \textit{strength} of these fields as well as of the incident laser radiation. In what follows, we assume the scenario of head--to--head collisions between ions and laser beams, typical for the Gamma-Factory experiments \cite{BuC20}. Let $\vec{v}$ be the velocity of ions in the laboratory frame. Then the corresponding Lorentz factor is $\gamma=\sqrt{1-\left(v/c\right)^2}$, where $c$ is the speed of light. The energy of an incident photon with frequency $\omega_{\rm lab}$ in the laboratory rest frame is boosted by the transition to the ion rest frame:
%
%
\begin{equation}
    \label{eq:energy}
    \hbar \omega_0 = 2 \gamma \hbar \omega_{\rm lab} \, . 
\end{equation}
%
%
Equation\,\eqref{eq:energy} implies that excitation energies of up to $\hbar \omega_0 \approx$\,40\,keV in the ion rest frame can be achieved for the typical parameters of the Gamma Factory (see Ref.\,\cite{BuC20}), when using commercially available lasers. One can obtain even higher energies by using a free-electron laser as a source of the primary radiation.

Additionally, we assume that moving ions are subject to a magnetic field $\mathbfcal{B}_\mathrm{lab}$ produced by the dipole magnets of the collider. This field is taken to be orthogonal to the ion velocity $\vec{v}$. In the ion rest frame, $\mathbfcal{B}_\mathrm{lab}$ is boosted by $\gamma$ and, at the same time, this magnetic field gives rise to a static electric field $\mathbfcal{E}$ through Lorentz transformation. In the SI units:
\begin{align}
    \mathbfcal{B} &= \gamma \mathbfcal{B}_\mathrm{lab}\ ,\label{eq:gammaB}\\
    \mathbfcal{E} &= \gamma \left[\vec{v}\times\mathbfcal{B}_\mathrm{lab}\right]\ .\label{eq:gammaE}
\end{align}
Within the ion rest frame, electric and magnetic fields are chosen to be along $x$ and $z$ axes, so that the vectors $\mathbfcal{E}$, $\mathbfcal{B}$ and the photon wave--vector ${\bm k}$ form a triplet of mutually orthogonal vectors, see Fig.\,\ref{Fig:geometry}a. With such an ``orthogonal'' configuration, $\mathbfcal{E} \perp \mathbfcal{B} \perp {\bm k}$, the APV dichroism resulting from the PV-Stark interference corresponds to the P-odd,T-even rotational invariant ${\mathcal {\bm E}} \times {\mathcal {\bm B}} \cdot {\bm \sigma}$, in principle, allowing a complete set of reversals, which is a crucial feature that can be used to separate the APV effect from systematics. Here ${\bm \sigma}||{\bm k}$ is the circular polarization of the incident light. The helicity of the light beam is $\lambda={\bm \sigma}\cdot \hat{{\bm k}}$.

Not only the spatial configuration of the fields, but also their \textit{strengths} are crucial for the investigations of parity violation in highly charged ions. In the proposed Gamma Factory experiments at the LHC, the magnetic-field strength of dipole magnets is $\mathcal{B}_\mathrm{lab}\approx 8$\,T. Hence the ions, moving with a relativistic Lorentz factor approaching $\gamma\approx 3000$, can experience an electric-field strength up to ${\mathcal E} \approx 7 \times 10^{10}$\,V/cm [see Eq.~\eqref{eq:gammaE}], which is much higher than typical laboratory field strengths \cite{Kra05}. The relativistic ions, circulating in the LHC ring, also ``see'' the boosted magnetic field $\mathcal{B}=\gamma\mathcal{B}_\mathrm{lab}\approx 2.4\times 10^4$~T [as in Eq.~\eqref{eq:gammaB}]. This field is large enough to resolve Zeeman sublevels of the ground and excited states of medium--$Z$ ions: in our case, the Zeeman splitting is of the order of $\mu_\mathrm{B}\mathcal{B}\approx 1.3$~eV, which should be compared with the energy-level widths in Table\,\ref{tab1} and the linewidth of the incident light in the ion rest frame.

\section{Theoretical background}
\label{sec:theory}

Having specified the geometry and parameters of the PV study with highly charged ions at the Gamma Factory, we are ready to review its basic theory. In what follows, we remind how to account for the mixing of opposite--parity ionic levels induced by the weak electron--nucleus interaction as well as by external electric field. Moreover, the evaluation of the matrix elements for the multipole (electric and magnetic) radiative transitions is briefly discussed at the end of this Section. 

\subsection{Weak interaction mixing}
\label{subsection:pv_mixing}

One of the important consequences of the Glashow--Weinberg--Salam electroweak theory is the prediction that, apart from the electromagnetic interaction, there is also the weak interaction between electrons and nucleus in an atom (ion). The dominant nuclear--spin--independent part of this interaction is mediated by exchange of the $Z^0$ boson and is described by the effective Hamiltonian \cite{Khr91,BuK08,SaB18}:
\begin{equation}
    \label{eq:PV_Hamiltonian}
    \hat{H}_w({\bm r}) = -\frac{G_F}{\sqrt{8}} \, Q_w \, \rho_N({\bm r})\, \gamma_5 \, ,
\end{equation}
where $G_F$ is the Fermi coupling constant, $Q_w \approx -N + Z \left(1 - 4 \sin^2\theta_W\right)$ is the weak charge of the nucleus which consists of $Z$ protons and $N$ neutrons, 
$\gamma_5$ is the ``fifth'' Dirac matrix, and $\rho_N({\bm r})$ is the nuclear weak--charge density normalized to unity:
\begin{equation}
  \int \rho_N({\bm r}) d^3r=1\ .
\end{equation}
The weak electron--nucleus interaction (\ref{eq:PV_Hamiltonian}) results in the \textit{mixing} of atomic or ionic states of opposite parity. To explore this mixing, one usually treats the Hamiltonian \eqref{eq:PV_Hamiltonian} within the first--order perturbation theory. The perturbation approach suggests that the wavefunctions of two closely--lying magnetic substates $\ketm{\alpha_a J_a P_a M_a}$ and $\ketm{\alpha_b J_b P_b M_b}$ with $P_a = - P_b$ are modified as:
\begin{align}
    \label{eq:state_PV_mixing}
    \Psi_{\alpha_a J_a M_a}({\bm \xi}) &\approx  \Psi^{(0)}_{\alpha_a J_a P_a M_a}({\bm \xi}) +i \eta_{w} \Psi^{(0)}_{\alpha_b J_b P_b M_b}({\bm \xi}) \, , \nonumber \\[0.2cm]
    \Psi_{\alpha_b J_b M_b}({\bm \xi}) &\approx  \Psi^{(0)}_{\alpha_b J_b P_b M_b}({\bm \xi}) +i \eta_{w} \Psi^{(0)}_{\alpha_a J_a P_a M_a}({\bm \xi}) \, ,
\end{align}
where $\Psi^{(0)}$ are solutions of unperturbed (purely electromagnetic) Hamiltonian and the short--hand notation ${\bm \xi} = \{ {\bm r}_1, {\bm r}_2, ... {\bm r}_N$ \} is used. Moreover, the mixing coefficient is given by:
\begin{align}
    \label{eq:pv_mixing_coefficient_general}
    \eta_{w} &= -i \, \delta_{J_a, J_b} \delta_{M_a, M_b} \delta_{P_a, -P_b} \nonumber \\[0.2cm]
    &\times \frac{\mem{\alpha_b J_b P_b M_b}{\sum_q \hat{H}_w({\bm r}_q)}{\alpha_a J_a P_a M_a}}{E_a - E_b - i\left(\Gamma_a-\Gamma_b\right)/2} \, ,
\end{align}
with $E_a$ and $E_b$ being the energies of the levels and $\Gamma_a$ and $\Gamma_b$ are their natural widths. As seen from Eq.\,\eqref{eq:pv_mixing_coefficient_general} the weak interaction leads to the mixing of ionic substates with the same total angular momenta $J_a = J_b$ and their projections $M_a = M_b$ but with opposite parities, $P_a = - P_b$. Note that the matrix elements of $H_w$ must be purely imaginary if one is to demand T-invariance \cite{Khr91}. 
The coefficient $\eta_{w}$ can be significantly enhanced if the energy splitting $E_a - E_b$ decreases. Such a ``tuning'' of energy spectra can be achieved by varying electronic configuration, nuclear charge $Z$, or even mass number $N + Z$  of partially stripped ions \cite{FeA10}.   

\subsection{Stark mixing}
\label{subsection:Stark_mixing}

Besides the weak force, the interaction of a partially stripped ion with an external electric field may also lead to mixing of opposite--parity states. For the case of a constant electric field ${\cal{\bm E}}$, this interaction is described by the operator:
\begin{equation}
    \label{eq:Stark_operator}
    \hat{H}_S({\bm r}) = - {\cal {\bm E}} I_4 \sum_q {\bm d}_q = e {\cal {\bm E}} I_4 \sum_q {\bm r}_q \, ,
\end{equation}
where ${\bm d}_q$ is the electric-dipole operator for a $q$--th electron and $I_4$ is 4$\times$4 unity matrix. By treating the operator $\hat{H}_S$ as a perturbation, we can find that the wavefunction of a particular ionic substate is modified by the Stark effect as:
\begin{equation}
    \label{eq:state_Stark_mixing}
    \Psi_{\alpha_a J_a M_a}({\bm \xi}) \approx  \Psi^{(0)}_{\alpha_a J_a P_a M_a}({\bm \xi})
    + \sum\limits_{J_b M_b} \eta_{S} \, \Psi^{(0)}_{\alpha_b J_b P_b M_b}({\bm \xi}) \, ,
\end{equation}
where the mixing coefficient is given by:
\begin{eqnarray}
    \label{eq:Stark_mixing_coefficient_general}
    \eta_{S} &=& \delta_{P_a, -P_b}  \frac{\mem{\alpha_b J_b P_b M_b}{\sum_q e {\bm r}_q {\cal {\bm E}} I_4}{\alpha_a J_a P_a M_a}}{E_a - E_b - i\left(\Gamma_a-\Gamma_b\right)/2} \, .
\end{eqnarray}
Depending on the mutual orientation of ${\cal {\bm E}}$ and the magnetic field ${\cal {\bm B}}$ chosen as the quantization axis, this coefficient can be non--zero for states with opposite parity as well as with $J_a - J_b = 0, \pm 1$ and $M_a - M_b = 0, \pm 1$. Similar to the weak--mixing coefficient $\eta_w$, the Stark-mixing coefficient $\eta_S$ is inversely proportional to the energy splitting $E_a - E_b$ and, hence, is enhanced for nearly degenerated levels. Below we discuss how the Stark effect can be used to control the state mixing and to extract its weak--interaction part. 
\subsection{Photoexcitation matrix element}
\label{subsection:transition_matrix_element}

As mentioned in Sec.\,\ref{section:setup}, the information about the (weak--interaction as well as Stark) mixing of ionic states of opposite parity can be obtained from the analysis of the excitation of ions by \textit{circularly} polarized light. All the properties of this process can be theoretically studied based on the matrix element:
\begin{equation}
    \label{eq:transition_matrix_element_general}
    {\cal M}^{(0)}_{M_i M_f} = \mem{\alpha_f J_f P_f M_f}{{\hat {\bm R}}({\bm k}, \, {\bm \epsilon}_\lambda)}{\alpha_i J_i P_i M_i} \, ,
\end{equation}
which describes transitions between (many--electron) substates $\ketm{\alpha_i J_i P_i M_i} \to \ketm{\alpha_f J_f P_f M_f}$ under absorption of a photon with the wave-- and polarization vectors ${\bm k}$ and ${\bm \epsilon}_\lambda$, respectively. In what follows, we briefly lay out the evaluation of the matrix element (\ref{eq:transition_matrix_element_general}) for the solutions of an \textit{unperturbed} Hamiltonian, which are characterized by well--defined parities $P_i$ and $P_f$.

To proceed with the analysis of the matrix element \eqref{eq:transition_matrix_element_general}, one has to define the explicit form of the photon-absorption operator ${\hat {\bm R}}({\bm k}, \, {\bm \epsilon}_\lambda)$. If the electron--photon coupling is treated in the Coulomb gauge, this operator can be written as:
\begin{equation}
    \label{eq:R_operator_general}
    {\hat {\bm R}}({\bm k}, \, {\bm \epsilon}_\lambda) = \sum\limits_q {\bm \alpha}_q \, {\bm \epsilon}_{\lambda} \, {\rm e}^{i {\bm k} {\bm r}_q} \, ,
\end{equation}
where ${\bm \alpha}_q = \{ \alpha_{x,q}, \alpha_{y,q}, \alpha_{z,q} \}$ is the vector of Dirac matrices for $q$--th electron \cite{Ros57}. Moreover, we here attribute the right and left circular polarization of photons, ${\bm \epsilon}_{\lambda}$, to their helicity  $\lambda = \pm 1$, i.e. to the spin projection onto the propagation direction. 

As usual in atomic structure calculations, one can express the electron--photon interaction operator ${\hat {\bm R}}$ in terms of its electric and magnetic multipole components, see Refs.\,\cite{Ros57,MaM00,SeS21} for further details. By inserting this multipole expansion into Eq.\,(\ref{eq:transition_matrix_element_general}) and performing some angular momentum algebra one can write the photoexcitation matrix element as:
\begin{eqnarray}
    \label{eq:transition_matrix_element_expanded}
    {\cal M}^{(0)}_{M_i M_b} &=& i \sqrt{\frac{6 \pi}{2J_f + 1}} \, D^1_{M \lambda}(\hat{\bm k}) \,
    \sprm{J_i M_i \, L M}{J_f M_f} \nonumber \\[0.2cm]
    &\times& \left(a_{M1} + i\lambda a_{E1} \right) \, ,
\end{eqnarray}
where $D^1_{M \lambda}(\varphi_k, \theta_k)$ is the Wigner D--matrix that depends on the direction of propagation $\hat{\bm k} = \{\theta_k, \varphi_k \}$ of incident photon and $M = M_f - M_i$. Moreover, in Eq.\,\eqref{eq:transition_matrix_element_expanded} we restricted the multipole decomposition to the electric (E1) and magnetic (M1) dipole operators whose \textit{reduced} matrix elements are denoted as:
\begin{eqnarray}
    \label{eq:E1_M1_reduced_matrix_elements}
    a_{E1} &=& \rmem{\alpha_f J_f}{{\bm \alpha} {\bm a}_{L=1}^{(e)}}{\alpha_i J_i} \, , \nonumber \\[0.2cm]
    a_{M1} &=& \rmem{\alpha_f J_f}{{\bm \alpha} {\bm a}_{L=1}^{(m)}}{\alpha_i J_i} \, .
\end{eqnarray}
Together with the weak- and Stark-mixing coefficients, $a_{E1}$ and $a_{M1}$ are the main ``building blocks'' to investigate the parity--violating phenomena in ions. Since the explicit form and properties of these matrix elements have been discussed extensively in the literature \cite{MaM00,SuF02}, we will not elaborate on them here. Instead, we just mention that $a_{E1}$ and $a_{M1}$ are directly related to the \textit{rates} for the E1 and M1 photoinduced transitions between fine--structure levels:
\begin{equation}
    \label{eq:excitation_rates}
    W_{E1, M1} = {\mathcal I} \, \frac{8 \pi^3 \alpha c^2}{\hbar \omega^2 \Delta\omega} \, \frac{1}{2J_a + 1} \, \left| a_{E1, M1}\right|^2 \, .
\end{equation}
Here ${\mathcal I}$, $\omega$ and $\Delta\omega$ are the intensity, angular frequency and frequency width of the incident laser radiation, as ``seen'' in the ion rest frame.

\section{Excitation of partially stripped ions}
\label{sec:results}

In the previous section we have briefly discussed the theoretical tools that can be used to explore how the weak interaction and Stark effect influence the photoexcitation of ions or atoms. The derived formulas are general and may be applied to any ion, independent of its nuclear charge, atomic mass and electronic structure. In what follows, we employ these expressions to analyze $1{\rm s}_{1/2} \to 2{\rm s}_{1/2}$ and $1{\rm s}^2 \; 2{\rm s}_{1/2} \to 1{\rm s}^2 \; 3{\rm s}_{1/2}$ transitions in hydrogen-- and lithium--like ions, which are of particular interest for the future Gamma Factory experiments. 

\subsection{Hydrogen--like ions}
\label{subsec:hlike_ions}

We start our analysis from the 1s$_{1/2} \to$~2s$_{1/2}$ laser--induced transition in hydrogen--like ions. Due to the fact that 2s$_{1/2}$ and 2p$_{1/2}$ excited states are energetically close to each other, see Table~\ref{tab1}, they are relatively strongly mixed by both the weak (\ref{eq:PV_Hamiltonian}) and Stark (\ref{eq:Stark_operator}) interactions. For the wavefunctions of the magnetic substates $\ketm{{\rm 2s}_{1/2} \, \mu_f = \pm 1/2}$ this mixing implies that:
\begin{eqnarray}
    \label{eq:wave_functions_2s}
    \psi_{2\rm{s}_{1/2} \mu_f}({\bm r}) &\approx&  \psi^{(0)}_{2\rm{s}_{1/2} \mu_f}({\bm r})
    + i \eta_{w} \psi^{(0)}_{2\rm{p}_{1/2} \mu_f}({\bm r}) \nonumber \\[0.2cm] 
    &+& \sum\limits_{\mu'_f} \eta_{S}(\mu_f, \mu'_f) \, \psi^{(0)}_{2\rm{p}_{1/2} \mu'_f}({\bm r}) \, ,
\end{eqnarray}
where the general expressions (\ref{eq:state_PV_mixing}) and (\ref{eq:state_Stark_mixing}) have been applied to the single--electron system. 

The weak and Stark mixing coefficients $\eta_{w}$ and $\eta_{S}$ from Eq.~(\ref{eq:wave_functions_2s}) can be calculated with high accuracy by employing the standard bound--state solutions of the Dirac equation:
\begin{equation}
    \label{eq:wave_function_hydrogenic}
    \psi^{(0)}_{n \kappa m}({\bm r}) = 
    \left( 
    \begin{array}{c}
        f_{n \kappa}(r) \Omega_{\kappa \mu}(\hat{\bm r})  \\[0.1cm]
        i g_{n \kappa}(r) \Omega_{-\kappa \mu}(\hat{\bm r})  
   \end{array}
    \right) \, ,
\end{equation}
which are characterized by the principal $n$ and Dirac $\kappa$ quantum numbers and where $m$ is the projection of the total angular momentum $j = |\kappa| - 1/2$. 
Here $\Omega_{\pm \kappa \mu}(\hat{\bm r})$ are the spin--angular functions, and $f_{n \kappa}(r)$, $g_{n \kappa}(r)$ are the ``large'' and ``small'' radial components. To generate these radial components we have taken into account the effect of the finite size of the nucleus, which is described by the two-parameter Fermi model \cite{Pa92}. 

By employing the wavefunction (\ref{eq:wave_function_hydrogenic}) in Eq.~(\ref{eq:pv_mixing_coefficient_general}) we can obtain the coefficient of the weak--interaction mixing between the hydrogenic states $\ketm{2{\rm s}_{1/2}}$ and $\ketm{2{\rm p}_{1/2}}$:
\begin{eqnarray}
    \label{eq:pv_mixing_coefficient_hydrogen}
    \eta_w &=& \frac{G_F \, Q_w}{\sqrt{8}} \, \frac{1}{E_{2\rm{s}_{1/2}} - E_{2\rm{p}_{1/2}} - i\left(\Gamma_{2\rm{s}_{1/2}}-\Gamma_{2\rm{p}_{1/2}}\right)/2} \nonumber\\[0.2cm]
    &\times& \int_0^\infty {\rm d}r \, r^2 \, \rho_N(r) \Bigg[g_{2\rm{p}_{1/2}}(r)f_{2\rm{s}_{1/2}}(r) 
    \nonumber \\[0.2cm]
    &-& g_{2\rm{s}_{1/2}}(r) f_{2\rm{p}_{1/2}}(r) \Bigg] \, .
\end{eqnarray}
Owing to the natural width term $-i\Gamma/2$ in the denominator, this coefficient is complex and its real and imaginary parts are related to each other as:
\begin{equation}
    \label{eq:pv_mixing_coefficient_Re_Im_relation}
    \frac{{\rm Re} \, \left(\eta_w\right)}{{\rm Im} \, \left(\eta_w\right)} = 2 \frac{E_{2\rm{s}_{1/2}} - E_{2\rm{p}_{1/2}}}{\Gamma_{2\rm{s}_{1/2}}-\Gamma_{2\rm{p}_{1/2}}} \, ,
\end{equation}
where $\left(\Gamma_{2\rm{s}_{1/2}}-\Gamma_{2\rm{p}_{1/2}}\right) \approx -\Gamma_{2\rm{p}_{1/2}}$ for the case of hydrogen--like ions. 

By making use of Eq.\,(\ref{eq:pv_mixing_coefficient_hydrogen}) the 2s$_{1/2}$--2p$_{1/2}$ (weak--interaction) mixing coefficient was calculated for the entire isoelectronic sequence. For the sake of brevity, we present in Table\,\ref{tab2} only its real part and recall that ${\rm Im} \, \eta_w$ can be trivially obtained from Eq.\,(\ref{eq:pv_mixing_coefficient_Re_Im_relation}). \asc{As seen from the table, $\eta_w$ grows linearly with the nuclear charge $Z$ in the low--$Z$ domain and even faster for medium-- and high--$Z$ ions. In order to understand this behaviour, we recall that in the nonrelativistic limit (i.e. for low--$Z$ ions), the effective weak--interaction Hamiltonian can be written as $\hat{H}_w \propto Q_w \left({\bm \sigma}{\bm p} \delta^{(3)}({\bm r}) + \delta^{(3)}({\bm r}) {\bm \sigma}{\bm p} \right)$, see Refs.~\cite{Khr91,BoB97,BuK08} for further details. Since both the weak charge $Q_w$ and linear momentum ${\bm p}$ scale linearly with $Z$ and $\left| \psi_{\rm s}(0) \right|^2 \propto Z^3$, the matrix element of the operator $\hat{H}_w$, written in the Pauli approximation, behaves as $\mem{\psi_{\rm s}}{\hat{H}_w}{\psi_{\rm p}} \propto Z^5$. Together with the Lamb shift 2s--2p$_{1/2}$--splitting $\Delta E \propto Z^4$, it implies $\eta_w \propto Z$ in the low--$Z$ domain. With the increase of the nuclear charge $Z$, however, the relativistic corrections to the matrix element of the weak--interaction Hamiltonian as well as higher--order terms in the Lamb shift \cite{YeP19}, result in quadratic (or even higher) dependence of the mixing $\eta_w$ parameter on $Z$.}  

%
%
\begin{table*}[tb] 
\begin{center}
\begin{tabular}{lllllll}
   \hline\\[-0.3cm]
    Ion \hspace*{1.2cm} & $E_{2\rm{s}} - E_{2\rm{p}}$ (eV) \hspace*{0.4cm}& $E_{2\rm{s}} - E_{1\rm{s}}$ (eV) \hspace*{0.4cm} & $\Gamma_{2\rm{p}}$ (eV)   \hspace*{0.6cm} & $\Gamma_{2\rm{s,M1}}$ (eV) \hspace*{0.5cm} & $\Gamma_{2\rm{s,2E1}}$ (eV) \hspace*{0.4cm} & $\Gamma_{2\rm{s,tot}}$ (eV) \\[0.2cm]
   \hline
   $^{1}$H  &  $4.37 \times 10^{-6}$  & $10.20$  &$4.12 \times 10^{-7}$& $1.64 \times 10^{-21}$  & $5.42 \times 10^{-15}$ & $5.42 \times 10^{-15}$ \\
   $^{20}$Ne$^{9+}$  &  $2.01 \times 10^{-2}$  & $1.02 \times 10^3$  &$4.13 \times 10^{-3}$& $1.65 \times 10^{-11}$  & $5.40 \times 10^{-9}$ & $5.41 \times 10^{-9}$ \\
   $^{40}$Ca$^{19+}$  &  $0.23$  & $4.10 \times 10^3$  &$6.62 \times 10^{-2}$& $1.72 \times 10^{-8}$  & $3.42 \times 10^{-7}$ & $ 3.60 \times 10^{-7}$ \\
   $^{64}$Zn$^{29+}$  &  $0.92$  & $9.28 \times 10^3$  &$0.34$& $1.02 \times 10^{-6}$  & $3.83 \times 10^{-6}$ &  $4.85 \times 10^{-6}$ \\
   $^{90}$Zr$^{39+}$  &  $2.48$  & $1.66 \times 10^4$  &$1.07$& $1.89 \times 10^{-5}$  & $2.11 \times 10^{-5}$ & $3.40 \times 10^{-5}$ \\
   $^{120}$Sn$^{49+}$  &  $5.40$  & $2.63 \times 10^4$  &$2.62$& $1.86 \times 10^{-4}$ & $7.81 \times 10^{-5} $ & $2.64 \times 10^{-4}$ \\
   $^{142}$Nd$^{59+}$  &  $10.45$  & $3.84 \times 10^4$  &$5.47$& $1.24 \times 10^{-3}$  & $2.26 \times 10^{-4}$ & $1.46 \times 10^{-3}$ \\
   $^{174}$Yb$^{69+}$  &  $19.19$  & $5.33 \times 10^4$  &$10.22$& $6.31 \times 10^{-3}$  & $5.47 \times 10^{-4}$ & $6.85 \times 10^{-3}$ \\
   $^{202}$Hg$^{79+}$  &  $34.56$  & $7.13 \times 10^4$  &$17.59$& $2.67 \times 10^{-2}$  & $1.16 \times 10^{-3}$ & $2.78 \times 10^{-2}$ \\
   $^{208}$Pb$^{81+}$  &  $39.02$  & $7.53 \times 10^4$  &$19.45$& $3.50 \times 10^{-2}$  & $1.34 \times 10^{-3}$ & $3.63 \times 10^{-2}$ \\
   $^{232}$Th$^{89+}$  &  $65.99$  & $9.29 \times 10^4$  &$28.44$& $9.87 \times 10^{-2}$  & $2.24 \times 10^{-3}$ & $0.10$ \\
   \hline
\end{tabular}
\caption{The energy splittings between $2{\rm s}$ and $2{\rm p}_{1/2}$ states (second column) as well as between 2s and 1s states (third column), see Ref.~\cite{Yer15} for both. Moreover, the natural widths of the $2{\rm p}_{1/2}$ state from Ref.~\cite{Jit05} (fourth column) \asc{and the $2{\rm s}_{1/2}$ state are displayed. The width of the $2{\rm s}_{1/2}$ state is divided into the contribution of the M1 decay (fifth column), the contribution of the two-photon E1 decay (sixth column) and the total width (last column). The values in last three columns are found in Ref.~\cite{Sa98}.}}
\label{tab1}
\end{center}
\end{table*}

Beside the weak--interaction coefficient (\ref{eq:pv_mixing_coefficient_hydrogen}), we have to analyze also the Stark--induced mixing. For the $\ketm{2{\rm s}_{1/2} \, \mu_f}$ and $\ketm{2{\rm p}_{1/2} \, \mu'_f}$ substates of a hydrogen--like ion, the mixing coefficient (\ref{eq:Stark_mixing_coefficient_general}) can be written as:
\begin{eqnarray}
    \label{eq:Stark_mixing_coefficient_hydrogen}
    \eta_S(\mu_f, \mu'_f) = \mathcal{E} \, &\sum\limits_{\mu}& \Big[ (-1)^\mu \, Y_{1 -\mu}(\theta_E, 0) \nonumber \\[0.2cm] 
    &\times& \sprm{j_a \mu_f \, 1\mu}{j_b \mu'_f} \, \tilde{\eta}_S \Big] \, ,
\end{eqnarray}
where $j_a = j_b = 1/2$ and the ``reduced'' counterpart $\tilde{\eta}_S$ is dependent neither on the strength nor on the direction of the external electric field:
\begin{eqnarray}
    \label{eq:Stark_mixing_coefficient_hydrogen_reduced}
    \tilde{\eta}_S &=& \sqrt{\frac{4\pi}{3}} \, 
    \frac{e \, [j_a]^{1/2} \, (-1)^{j_a-1/2} \, }{E_{2\rm{s}_{1/2}} - E_{2\rm{p}_{1/2}} - i\left(\Gamma_{2\rm{s}_{1/2}}-\Gamma_{2\rm{p}_{1/2}}\right)/2}\nonumber \\[0.2cm] 
    &\times& \Bigg[ 
    [l_a]^{1/2} \, (-1)^{l_b} \, \sprm{l_a 0 1 0}{l_b 0} \,
    \sixjm{l_a}{1/2}{j_a}{j_b}{1}{l_b} \nonumber \\[0.2cm] 
    &\times& \int_0^\infty {\rm d}r \, r^2 \, f_{2\rm{s}_{1/2}}(r)f_{2\rm{p}_{1/2}}(r) \nonumber \\[0.2cm]
    &+&  
    [l'_a]^{1/2} \, (-1)^{l'_b} \, \sprm{l'_a 0 1 0}{l'_b 0} \,
    \sixjm{l'_a}{1/2}{j_a}{j_b}{1}{l'_b} \nonumber \\[0.2cm] 
    &\times& \int_0^\infty {\rm d}r \, r^2 \, g_{2\rm{s}_{1/2}}(r) g_{2\rm{p}_{1/2}}(r) \Bigg] \, .
\end{eqnarray}
Here, $l_a = 0$, $l_b = 1$ and $l'_a = 1$, $l'_b = 0$ are the orbital angular momenta of the large and small components of the Dirac wavefunctions $\psi^{(0)}_{2\rm{s}_{1/2} \mu}({\bm r})$ and $\psi^{(0)}_{2\rm{p}_{1/2} \mu}({\bm r})$
, and the short--hand notation $[j] = 2j + 1$ is introduced. 

Similar to $\eta_w$, the ``reduced'' mixing coefficient $\tilde{\eta}_S$ is also complex--valued and its real and complex parts are related to each other by an equation similar to Eq.~(\ref{eq:pv_mixing_coefficient_Re_Im_relation}). In Table~\ref{tab2}, therefore, we display only ${\rm Re}(\tilde{\eta}_S)$ that was calculated for the 2s$_{1/2}$--2p$_{1/2}$ mixing in various hydrogen--like ions. In contrast to the $\eta_w$, this Stark coefficient decreases with the increase of nuclear charge $Z$. \asc{The most pronounced decline in the Stark--mixing parameter can be observed in the low--$Z$ regime, where $\eta_S \propto Z^{-5}$. Similar to the weak--interaction case, this behaviour can be understood from the $Z$--scaling of the 2s--2p$_{1/2}$ energy splitting, $\Delta E \propto Z^4$, as well as of the electric dipole matrix element $\mem{\psi_{\rm s}}{e{\bm r}}{\psi_{\rm p}} \propto 1/Z$.} 

Having evaluated the weak--interaction and Stark 2s--2p--mixing coefficients, we are ready to further discuss the 1s$_{1/2} \to$~2s$_{1/2}$ laser excitation of hydrogen--like ions. The matrix element for this transition:
\begin{eqnarray}
    \label{eq:transition_matrix_element_1s-2s}
    {\cal M}_{\mu_i \mu_f}({\rm 1s-2s}) &=& {\cal M}^{(0)}_{\mu_i \mu_f}({\rm 1s-2s}; \, M1) \nonumber \\[0.2cm]
    && \hspace*{-1cm} - i \eta^*_{w} \, {\cal M}^{(0)}_{\mu_i \mu_f}({\rm 1s-2p}; \, E1) \nonumber \\[0.2cm]
    && \hspace*{-1cm} + \sum\limits_{\mu'_f} \eta^*_S(\mu_f, \mu'_f)  \, {\cal M}^{(0)}_{\mu_i \mu'_f}({\rm 1s-2p}; \, E1)
\end{eqnarray}
is obtained with the help of the wavefunction (\ref{eq:wave_functions_2s}). Here, ${\cal M}^{(0)}_{\mu_i \mu_f}({\rm 1s-2s}; \, M1)$ and ${\cal M}^{(0)}_{\mu_i \mu_f}({\rm 1s-2p}; \, E1)$ describe the magnetic and electric dipole photoexcitation to the \textit{unperturbed} 2s$_{1/2}$ and 2p$_{1/2}$ excited states. By making use of Eqs.~(\ref{eq:transition_matrix_element_expanded}) and (\ref{eq:Stark_mixing_coefficient_hydrogen}) we can further evaluate the transition matrix element as:
\begin{eqnarray}
    \label{eq:transition_matrix_element_1s-2s_final}
    {\cal M}_{\mu_i \mu_f}({\rm 1s-2s}) &=& \frac{i}{4 \pi} \left(a_{M1} + \lambda \eta^*_w a_{E1} \right) \, \mathcal{D}^{0}_{\mu_i \mu_f \lambda}
    \nonumber \\[0.2cm]
    &-& \lambda \, \mathcal{E} \, \tilde{\eta}^*_S \, a_{E1} \, 
    \mathcal{D}^{1}_{\mu_i \mu_f \lambda} \, ,
\end{eqnarray}
and to express it in terms of the reduced counterparts $a_{E1}$ and $a_{M1}$, and the mixing parameters $\eta_w$ and $\tilde{\eta}_S$. Moreover, ${\cal M}_{\mu_i \mu_f}({\rm 1s-2s})$ still depends on the direction of propagation $\left(\theta_k, \varphi_k \right)$ and helicity $\lambda$ of incident photons and the orientation of the electric field, $\theta_E$. In Eq.\,(\ref{eq:transition_matrix_element_1s-2s_final}) we introduced the function:
\begin{eqnarray}
    \label{eq:D_function}
    \mathcal{D}^{\nu}_{\mu_i \mu_f \lambda} &\equiv& 
    \mathcal{D}^{\nu}_{\mu_i \mu_f \lambda}\left(\theta_k, \varphi_k, \theta_E \right) \nonumber \\[0.2cm]
    && \hspace*{-1cm} =
    \sqrt{3 \pi} \sum\limits_{M M'} \Bigg[ D^1_{M \lambda}\left(\varphi_k, \theta_k \right) \,
    Y_{1 \mu}(\theta_E, 0) \nonumber \\[0.0cm]
    && \hspace*{-1cm} \times \sprm{1/2 \mu_f \, \nu \mu}{1/2 M'} \, \sprm{1/2 \mu_i \, 1 M}{1/2 M'} \Bigg] \, ,
\end{eqnarray}
to account for these angular and polarization dependencies. The analysis of the function $\mathcal{D}^{\nu}_{\mu_i \mu_f \lambda}$ suggests that the weak--interaction-- and Stark--induced transition amplitudes are \textit{in--phase} if the electric field ${\mathcal {\bm E}}$ and the light propagation ${\bm k}$ directions are orthogonal to each other and to the quantization ${\mathcal {\bm B}}$--axis. This enables the weak--interaction---Stark interference, which opens a way for the detailed and controllable study of the parity--violating mixing of atomic levels.

By choosing the ``orthogonal'' geometry ${\mathcal {\bm E}} \perp {\mathcal {\bm B}} \perp {\bm k}$, displayed in the upper panel of Fig.~\ref{Fig:geometry}, and by assuming, moreover,  a reasonably strong magnetic field ${\mathcal B}$, we are ready to analyze transitions $\ketm{{\rm 1s} \, \mu_i} \to \ketm{{\rm 2s} \, \mu_f}$ between \textit{individual sublevels}, split by the Zeeman effect. With the help of the matrix element (\ref{eq:transition_matrix_element_1s-2s_final}) we can obtain the rate for these (magnetic--sublevel) transitions:
\begin{eqnarray}
    \label{eq:rate_1s_2s_magnetic_sublevel}
    W_{1\rm{s} \mu_i \, - \, 2\rm{s} \mu_f}(\lambda) &=& W_{M1} \left( 1+2\lambda\frac{a_{E1}}{a_{M1}}\operatorname{Re}\left(\eta_w\right) \right. \nonumber \\
    &+& \frac{\mathcal{E}^2}{4\pi} \frac{a_{E1}^2}{a_{M1}^2} \left|\tilde\eta_s\right|^2 \pm\frac{\mathcal{E}}{\sqrt{\pi}} \frac{a_{E1}}{a_{M1}} \operatorname{Re} \left(\tilde\eta_s\right) \nonumber \\
    &\pm& \left. \frac{\mathcal{E}}{\sqrt{\pi}}\lambda \frac{a_{E1}^2}{a_{M1}^2}\operatorname{Re}\left(\tilde{\eta}_s\eta_w^*\right) \right) ,
\end{eqnarray}
where we neglected the tiny $\eta^2_w$ term. The \textit{plus} and \textit{minus} signs correspond here to the transitions $\ketm{{\rm 1s} \, \mu_i} \to \ketm{{\rm 2s} \, \mu_f = -1/2}$ and $\ketm{{\rm 1s} \, \mu_i} \to \ketm{{\rm 2s} \, \mu_f = +1/2}$, respectively. 

\asc{The expression~(\ref{eq:rate_1s_2s_magnetic_sublevel}) is rather general and can be used to identify certain limiting cases known from the earlier studies with neutral atoms. For example, for the vanishing electric field, $\mathcal{E}=0$, only the first two terms survive and the transition rate reads as $W_{1\rm{s} \mu_i \, - \, 2\rm{s} \mu_f}(\lambda) = W_{M1} \, \left(1 +  2 \lambda \operatorname{Re}\left(\eta_w\right) \left(a_{E1}/a_{M1}\right) \right)$. This is like in the experiments on optical rotation in Bi, Tl, and Pb atoms, which were among the first observations of atomic parity violation, see Ref.~\cite{Khr91} for further details. The disadvantage of this scenario, however, is the absence of the electric field reversals, that were proven useful for controlling systematics. The experiments with highly forbidden $M1$ transitions as can be observed in neutral Cs and Yb atoms, provide the other limiting case. In this scenario, only the third and the fifth terms survive in the transition rate (\ref{eq:rate_1s_2s_magnetic_sublevel}) and the PV effect arises exclusively due to the interference of the Stark-- and PV--induced $E1$ amplitudes.}

\asc{As seen from Eq.~(\ref{eq:rate_1s_2s_magnetic_sublevel}), the transition rate} depends on the helicity $\lambda = \pm 1$ of incident light, thus implying that the probability to excite $\ketm{2{\rm s}_{1/2} \mu_f}$ sub--states is different for the absorption of left-- and right circularly polarized photons. Most naturally, the difference between the excitation rates $W(+1) \equiv W_{1\rm{s} \mu_i \, - \, 2\rm{s} \mu_f}(\lambda = + 1)$ and $W(-1) \equiv W_{1\rm{s} \mu_i \, - \, 2\rm{s} \mu_f}(\lambda = - 1)$ can be characterized by the \textit{circular dichroism} parameter:
\begin{equation}
    \label{eq:circular_dicroism_definition}
    \mathcal{A} = \frac{W(+1) - W(-1)}{W(+1) + W(-1)} \, .
\end{equation}
\asc{Of course, this parameter should be discussed together with the transition rate (\ref{eq:rate_1s_2s_magnetic_sublevel}). As we will see it later, the large values of $\mathcal{A}$ may correspond to almost vanishing rates $W_{1s \mu_i \, - \, 2s \mu_f}(\lambda)$, thus making experimental analysis of APV phenomena troublesome. For the future Gamma Factory experiments, therefore, it will be important to determine a parameter range for which both $\mathcal{A}$ and $W_{1s \mu_i \, - \, 2s \mu_f}(\lambda)$ are sufficiently large. Our calculations, presented below, will help to perform such an analysis and to guide future APV measurements.}

By inserting the rate (\ref{eq:rate_1s_2s_magnetic_sublevel}) into th expression (\ref{eq:circular_dicroism_definition}), we can obtain the dichroism for the laser--induced $\ketm{{\rm 1s} \, \mu_i} \to \ketm{{\rm 2s} \, \mu_f = \pm 1/2}$ transitions between (resolved) Zeeman sublevels:
\begin{eqnarray}
    \label{eq:dichroism_1s_2s_sublevels}
    \mathcal{A}_{\mu_i \mu_f} &\approx&  2 \, {\rm Re}(\eta_w) \, \frac{a_{E1}}{a_{M1}} \nonumber \\[0.2cm] 
    &\times& \frac{1 \pm \frac{\mathcal{E}}{\sqrt{4 \pi}}  \, \frac{a_{E1}}{a_{M1}} \, {\rm Re}(\tilde{\eta}_S)}{1 +  \frac{\mathcal{E}^2}{4 \pi} \, \left(\frac{a_{E1}}{a_{M1}}\right)^2 \, \left|\tilde{\eta}_S\right|^2 \pm \frac{\mathcal{E}}{\sqrt{\pi}} \, \frac{a_{E1}}{a_{M1}} \, {\rm Re}(\tilde{\eta}_S)} \, ,
\end{eqnarray}
where, again, \textit{plus} and \textit{minus} signs correspond to $\mu_f = \mp 1/2$, respectively. 

As seen from Eq.\,(\ref{eq:dichroism_1s_2s_sublevels}), the circular dichroism $\mathcal{A}_{\mu_i \mu_f}$ is directly proportional to the real part of mixing parameter $\eta_w$. One can obtain, therefore, the information about the weak interaction between electron and nucleus from the measurement of $\mathcal{A}_{\mu_i \mu_f}$ and based on the accurate calculations of the ratio of reduced matrix elements $a_{E1}/a_{M1}$. Moreover, the circular dichroism can be modified by an applied electric field $\mathcal{E}$, as follows from the second line of Eq.~(\ref{eq:dichroism_1s_2s_sublevels}). This $\mathcal{E}$--dependence of the $\mathcal{A}_{\mu_i \mu_f}$ can help to extract the weak--interaction parameter $\eta_w$ from the future experimental data in a controlled way.

\begin{figure}
    \centering
    \includegraphics[width=0.99\linewidth]{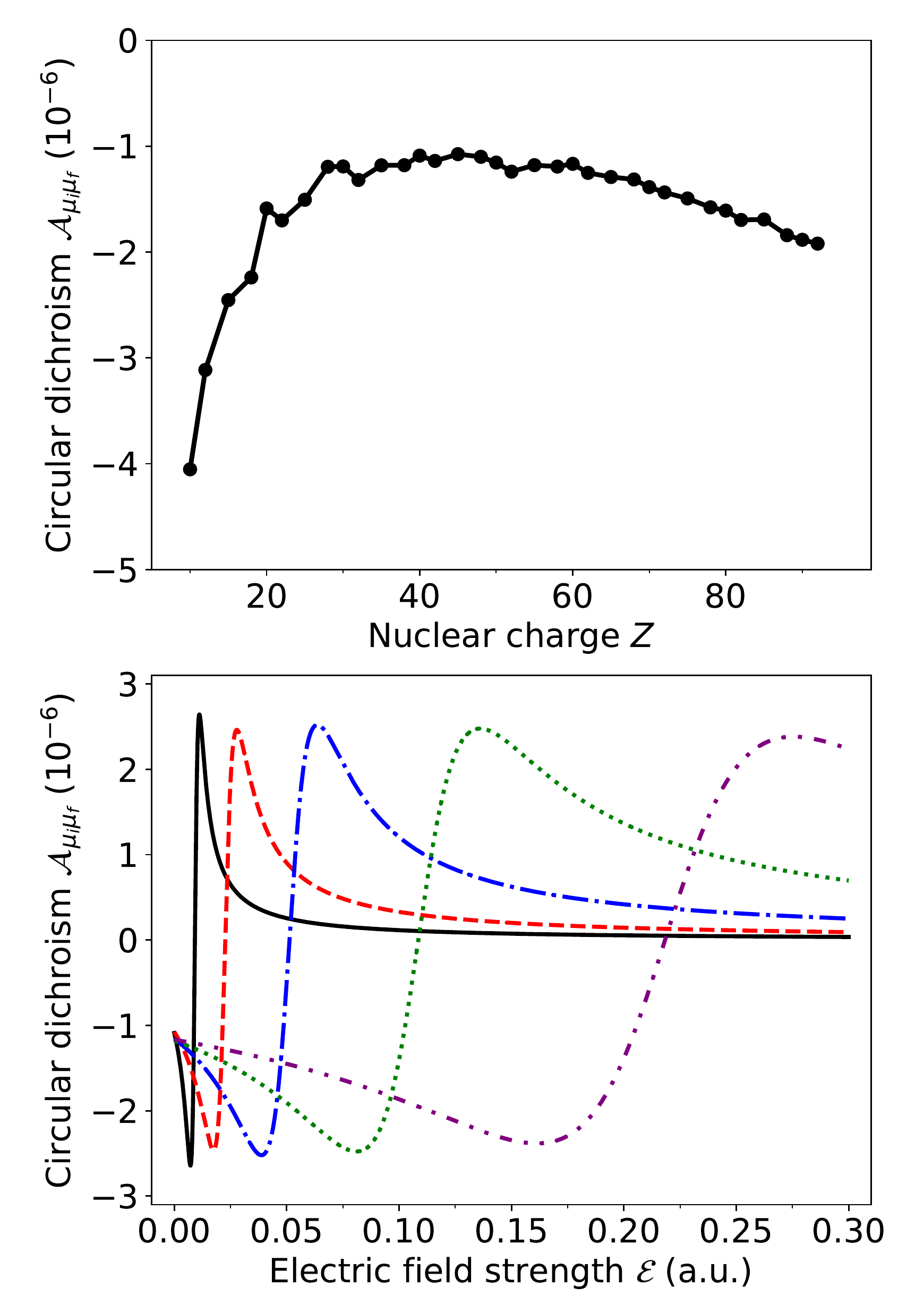}
    \caption{The circular dichroism (\ref{eq:dichroism_1s_2s_sublevels}) for the $\ketm{{\rm 1s}_{1/2} \, \mu_i} \to \ketm{{\rm 2s}_{1/2} \, \mu_f = +1/2}$
    transition in hydrogen--like ions. In the upper panel, the parameter $\mathcal{A}_{\mu_i \mu_f}$ is displayed as a function of the nuclear charge $Z$ and for \textit{zero} electric field, $\mathcal{E} = 0$.  In the lower panel, in contrast, we display the variation of the dicroism with $\mathcal{E}$ for $Z$~=~40 (black solid line), 45 (red dashed line), 50 (blue dash-dotted line), 55 (green dotted line) and 60 (purple dash-dot-dotted line).}
    \label{Fig2}
\end{figure}
\begin{figure}
    \centering
    \includegraphics[width=0.99\linewidth]{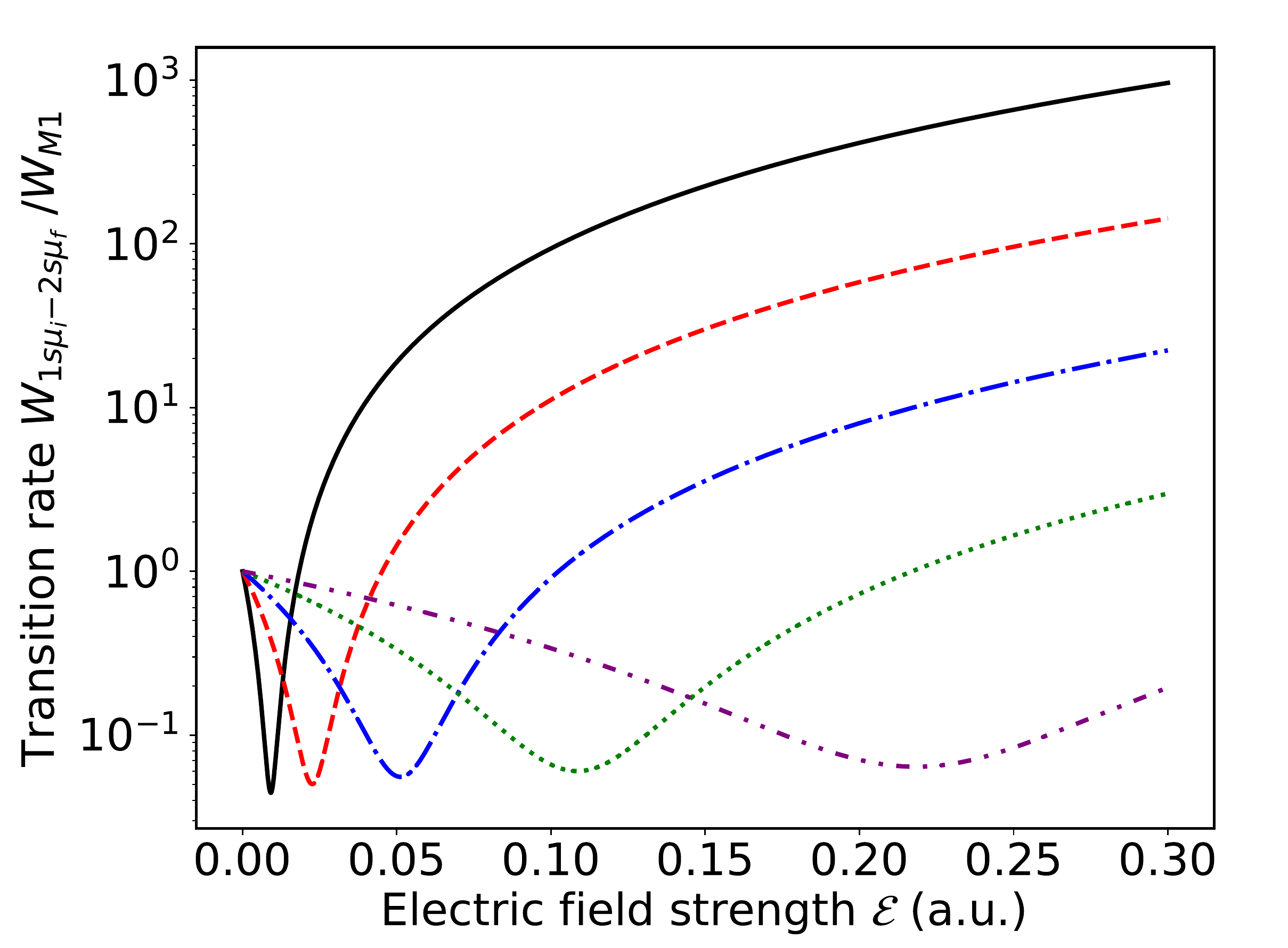}
    \caption{\asc{The transition rate  (\ref{eq:rate_1s_2s_magnetic_sublevel}) for the $\ketm{{\rm 1s}_{1/2} \, \mu_i} \to \ketm{{\rm 2s}_{1/2} \, \mu_f = +1/2}$
    transition in hydrogen--like ions. We display the variation of the transition rate with $\mathcal{E}$ for $Z$~=~40 (black solid line), 45 (red dashed line), 50 (blue dash-dotted line), 55 (green dotted line) and 60 (purple dash-dot-dotted line).}}
    \label{Fig:rate}
\end{figure}

In order to better understand the effect of an applied electric field, we first consider the case $\mathcal{E} = 0$, for which the circular dichroism $\mathcal{A}_{\mu_i \mu_f} \approx 2 \, {\rm Re}(\eta_w) \, a_{E1}/ a_{M1}$ depends solely on the nuclear charge $Z$ of an ion. As seen from the upper panel of Fig.\,\ref{Fig2}, the $\mathcal{A}_{\mu_i \mu_f}(\mathcal{E} = 0)$ remains on the order of 10$^{-6}$ over the entire isoelectronic sequence. This behaviour is expected from the different $Z$--scalings of the mixing parameter and the amplitude ratio \cite{Pal88}. We argue, therefore, that the parity violation studies with hydrogen--like ions not necessarily have to be performed in the high--$Z$ domain, see also \cite{ZoB97}. The 2s--2p level mixing, caused by the weak interaction, can be observed also for medium--$Z$ ions, for which $\mathcal{A}_{\mu_i \mu_f} \approx - 10^{-6}$. The ions with 20~$\lesssim Z \lesssim$~50 can be even more preferable for the Gamma Factory experiments in comparison to their heavier counterparts. This is due to the fact that the 1s$_{1/2} \to$~2s$_{1/2}$ transition in \textit{relativistic} medium--$Z$ ions can be induced by conventional (visible--light) laser, whose frequency is boosted to the EUV regime by the Doppler effect \cite{ZoB97}. Moreover, the widths of the relevant levels increase with Z (see Table\,\ref{tab1}), hence stronger magnetic fields are required to resolve the magnetic sublevels with Zeeman effect. As mentioned above (Sec.\,\ref{section:setup}), the characteristic field strength in Gamma Factory is $\mathcal{B}\approx 2.4\times 10^4$\,T, which can induce Zeeman splitting of the order of $\mu_\mathrm{B}\mathcal{B}\approx 1.3$\,eV. This is comparable to the 2p--level widths of the medium--$Z$ ions in Table\,\ref{tab1}.

\begin{table}[bt] 
\begin{tabular}{llll}
   \hline\\[-0.3cm]
    Ion \hspace*{0.6cm} &  $\operatorname{Re}(\eta_w)$  \hspace*{0.8cm} & $\operatorname{Re}(\tilde\eta_s)$ (a.u.) \hspace*{0.3cm} & $a_{E1}/a_{M1}$  \\[0.2cm]
   \hline
   $^{1}$H  &   $-4.61 \times 10^{-13}$  & $-3.81 \times 10^7$ & $-1.58 \times 10^7$ \\
   $^{20}$Ne$^{9+}$  & $1.28 \times 10^{-10}$  & $-8.21 \times 10^2$ & $-1.58 \times 10^4$ \\
   $^{40}$Ca$^{19+}$  & $4.03 \times 10^{-10}$  & $-35.61$ & $-1.96 \times 10^3$ \\
   $^{64}$Zn$^{29+}$  &  $1.03 \times 10^{-9}$  & $-5.74$ &  $-5.73 \times 10^2$ \\
   $^{90}$Zr$^{39+}$  & $2.24 \times 10^{-9}$  & $-1.55$ & $-2.37 \times 10^2$ \\
   $^{120}$Sn$^{49+}$  &$4.70 \times 10^{-9}$ & $-0.55 $ & $-1.19 \times 10^2$ \\
   $^{142}$Nd$^{59+}$  &  $8.33 \times 10^{-9}$  & $-0.23$ & $-66.46$ \\
   $^{174}$Yb$^{69+}$  & $1.60 \times 10^{-8}$  & $-0.10$ & $-40.20$ \\
   $^{202}$Hg$^{79+}$  & $2.90 \times 10^{-8}$  & $-4.81 \times 10^{-2}$ & $-25.61$ \\
   $^{208}$Pb$^{81+}$  &  $3.25 \times 10^{-8}$  & $-4.12 \times 10^{-2}$ & $-23.51$ \\
   $^{232}$Th$^{89+}$  &  $4.90 \times 10^{-8}$  & $-2.16 \times 10^{-2}$ & $-16.88$ \\
   \hline
\end{tabular}
\caption{Parameters relevant to parity--violation studies with $1{\rm s}_{1/2} \to 2{\rm s}_{1/2}$ photoexcitation of hydrogen--like ions. In particular, we present here the real parts of the weak--interaction mixing coefficient (\ref{eq:pv_mixing_coefficient_hydrogen}) and of the \textit{reduced} Stark mixing coefficient (\ref{eq:Stark_mixing_coefficient_hydrogen_reduced}). We note that imaginary parts of both coefficients can be trivially obtained based on Eq.~(\ref{eq:pv_mixing_coefficient_Re_Im_relation}). Finally, the ratio of the reduced matrix elements of the E1 ($1{\rm s} \to 2{\rm p}_{1/2}$) and M1 ($1{\rm s} \to 2{\rm s}$) transitions is given in the last column.}
\label{tab2}
\end{table}

Yet another advantage of medium--$Z$ ions (over high--$Z$ ones) is their greater sensitivity to an external electric field $\mathcal{E}$. This is seen, for example, from the large values of the ``reduced'' Stark mixing parameter $\tilde{\eta}_S$ for $Z \lesssim$\,50, displayed in Table\,\ref{tab2}. In the middle--$Z$ domain, therefore, even a moderate electric field, $|\mathcal{E}| \ll 1$~a.u., can remarkably modify the circular dichroism (\ref{eq:dichroism_1s_2s_sublevels}). In order to illustrate this, we display $\mathcal{A}_{\mu_i \mu_f}$ as a function of $\mathcal{E}$ in the lower panel of Fig.\,\ref{Fig2}. The calculations were performed for the $\ketm{{\rm 1s}_{1/2} \, \mu_i} \to \ketm{{\rm 2s}_{1/2} \, \mu_f = +1/2}$ transition in hydrogen--like Zr$^{39+}$ (black solid line), Rh$^{44+}$ (red dashed line), Sn$^{49+}$ (blue dash--dotted line), Cs$^{54+}$ (green dotted line) and Nd$^{59+}$ (purple dash--dot--dotted line) ions. As seen from the figure, the circular dichroism varies greatly, depending on the field strength $\mathcal{E}$, and even exhibits resonant--like behaviour. \asc{This behaviour is caused by a mutual cancellation of the parity--conserving terms in the denominator of Eq.\,(\ref{eq:dichroism_1s_2s_sublevels}). Indeed, the cancellation of the parity conserving terms and, hence, reduction of the denominator, leads to almost a \textit{factor of three} enhancement of the absolute value of the circular dichroism. However, this decrease also results in a smaller transition rate which is displayed in Fig. \ref{Fig:rate}.} Moreover, the parameter $\mathcal{A}_{\mu_i \mu_f = +1/2}$ even changes its sign at $\mathcal{E}_0 = \sqrt{4\pi} / \left(a_{E1}/a_{M1} \, {\rm Re}\left(\tilde{\eta}_S\right) \right)$. The value of this ``critical'' field $\mathcal{E}_0$ increases with the nuclear charge of an ion: while being just $\mathcal{E}_0 \approx 0.01$~a.u. for $Z = 40$, it raises to $\mathcal{E}_0 \approx 0.22$~a.u. for $Z = 60$. This behaviour can be well expected from the fact that heavier ions are more ``robust'' against variation of $\mathcal{E}$. Again, it stresses the importance of medium--$Z$ ions for studying the weak--interaction phenomena in atomic systems. The 2s--2p level mixing in these ions can be modified by applying moderate electric fields and can help to extract the coefficient $\eta_w$ with high accuracy.     

\begin{figure}
    \centering
    \includegraphics[width=0.99\linewidth]{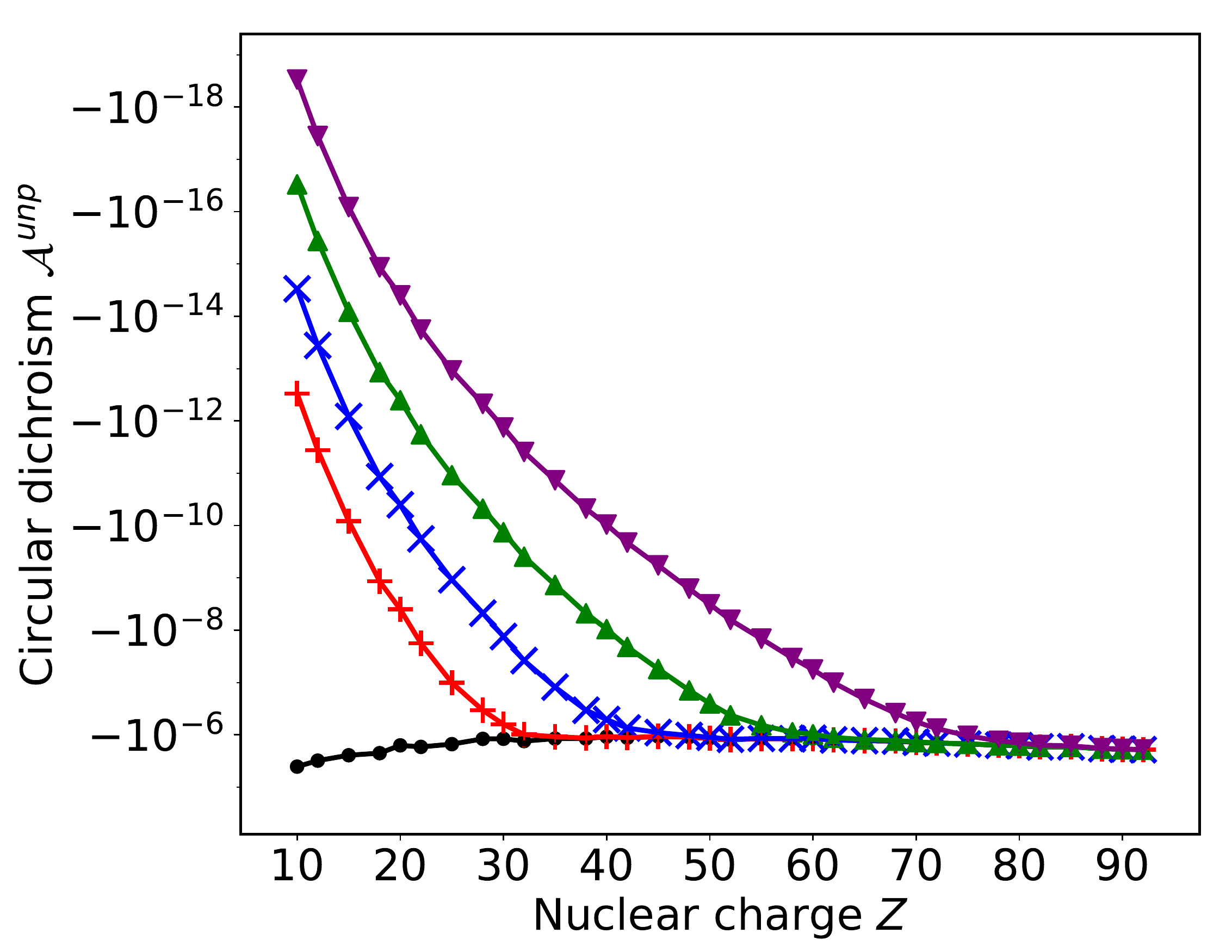}
    \caption{The circular dichroism parameter (\ref{eq:dichroism_1s_2s_unpolarized}) for the 1s$_{1/2} \to$~2s$_{1/2}$ transition in hydrogen--like ions. This ``unpolarized'' parameter is obtained for the case when magnetic sublevels of initial and final ionic states are unresolved. The calculations were performed for the external electric field with strength $\mathcal{E}$~=~0~a.u (black "$\bullet$"-line), 0.001~a.u. (red-"$+$" line), 0.01~a.u. (blue "x"-line), 0.1~a.u. (green "$\blacktriangle$"-line), and 1~a.u. (purple "$\blacktriangledown$"-line).}
    \label{Fig3}
\end{figure}

Until now, we discussed circular dichroism (\ref{eq:dichroism_1s_2s_sublevels}) for laser--induced transitions between particular magnetic sublevels, $\ketm{{\rm 1s}_{1/2} \, \mu_i} \to \ketm{{\rm 2s}_{1/2} \, \mu_f}$. Apart from this---preferred---experimental scenario, one can also consider the excitation of an unpolarized ion, whose Zeeman sublevels are \textit{not} resolved. The rate for this excitation is:
\begin{eqnarray}
    \label{eq:rate_1s_2s_unpolarized}
    W^{\rm (unp)}_{\rm{1s-2s}}(\lambda) &=& W_{M1} \left( 1 + 2 \lambda \frac{a_{E1}}{a_{M1}} {\rm Re}(\eta_w) + \frac{\mathcal{E}^2}{4\pi} \frac{a^2_{E1}}{a^2_{M1}} \left| \tilde{\eta}_S\right|^2 \right) \, ,
\end{eqnarray}
is obtained upon averaging (and summing) of the squared matrix element (\ref{eq:transition_matrix_element_1s-2s_final}) over the initial and final ionic substates.
As can be expected from the symmetry considerations, this ``unpolarized'' rate is independent on the directions of external electric and magnetic fields. It still depends, however, on the helicity $\lambda = \pm 1$ of the incident light, thus allowing us to obtain the circular dichroism: 
\begin{eqnarray}
    \label{eq:dichroism_1s_2s_unpolarized}
    \mathcal{A}^{\rm (unp)} &\approx&  2 \, {\rm Re}\left(\eta_w\right) \, \frac{a_{E1}}{a_{M1}} \, \frac{1}{1 + \frac{\left|\tilde{\eta}_s\right|^2 \mathcal{E}^2}{4\pi} \, \frac{a^2_{E1}}{a^2_{M1}}} \, .
\end{eqnarray}
Similar to $\mathcal{A}_{\mu_i \mu_f}$, it is directly proportional to the real part of the mixing parameter $\eta_w$, and can be also used to study the weak--interaction mixing of 2$s$ and 2$p$ levels. In contrast to the ``sublevel--resolved'' case (\ref{eq:dichroism_1s_2s_sublevels}), however, the ``unpolarized'' dichroism $\mathcal{A}^{\rm (unp)}$ is less sensitive to the external electric field. As seen from Eqs.\,(\ref{eq:rate_1s_2s_unpolarized})--(\ref{eq:dichroism_1s_2s_unpolarized}), no weak--interaction---Stark interference and, hence, no resonant behaviour of $\mathcal{A}^{\rm (unp)}$ can be observed after the summation (averaging) over Zeeman sublevels. Indeed, Fig.\,\ref{Fig3} shows that while $\mathcal{A}^{\rm (unp)} = \mathcal{A}_{\mu_i \mu_f} \approx - 10^{-6}$ for $\mathcal{E} = 0$, its absolute value steadily decreases with the growth of $\mathcal{E}$. This behavior arises from the fact that at zero electric field, the transition is M1 and the dichroism results from M1-PV interference (for which the effect averaged over Zeeman sublevels is nonzero). When electric field is applied, the transition rate is increased due to Stark mixing but this only dilutes the dichroism in the case of unresolved sublevels.


%
%
%
\subsection{Lithium--like ions}
\label{subsec:Lilike_ions}

As the second scenario for the parity--violation studies with partially striped ions at the Gamma Factory, we briefly discuss ${\rm 1s}^2 \; {\rm 2s_{1/2}} \to {\rm 1s}^2 \; {\rm 3s_{1/2}}$ photoexcitation of lithium--like species. For this case, \textit{both} initial and final ionic states are mixed with the closely--lying  ${\rm 1s}^2 \; {\rm 2p}_{1/2}$ and ${\rm 1s}^2 \; {\rm 3p}_{1/2}$ states of opposite parity. Once again, we apply the general theory from Sec.\,\ref{sec:theory} to estimate how this (weak--interaction and Stark) mixing affects the wavefunctions of magnetic sublevels:
\begin{eqnarray}
    \label{eq:wave_functions_2s_li_like}
    \Psi_{n \rm{S}_{1/2} M_f}({\bm \xi}) &\approx&  \Psi^{(0)}_{n \rm{S}_{1/2} M_f}({\bm \xi})
    + i \eta_{w} \Psi^{(0)}_{n \rm{P}_{1/2} M_f}({\bm \xi}) \nonumber \\[0.2cm] 
    &+& \sum\limits_{M'_f} \eta_{S} \, \Psi^{(0)}_{n \rm{P}_{1/2} M'_f}({\bm \xi}) \, ,
\end{eqnarray}
where $n = 2, 3$ and the short--hand notations $\ketm{{\rm 1s}^2 \; n{\rm s}: \, ^{2}{\rm S}_{1/2}} = \ketm{n \; ^{2}{\rm S}_{1/2}}$ and $\ketm{{\rm 1s}^2 \; n{\rm p}: \, ^{2}{\rm P}_{1/2}} = \ketm{n \; ^{2}{\rm P}_{1/2}}$ are used. Many--electron calculations are required in general to find the mixing coefficients $\eta_{w}$ and $\eta_{S}$ that enter Eq.~(\ref{eq:wave_functions_2s_li_like}). For medium-- and high--$Z$ lithium--like ions, however, these calculations can be performed with the help of the single--active electron approach with the frozen $K$--shell core. In this approach, the wavefunctions of a valence electron in the $n {\rm s}_{1/2}$ and $n {\rm p}_{1/2}$ states are obtained as solutions of a single--particle Dirac equation with an effective screening potential and substituted into Eqs.~(\ref{eq:pv_mixing_coefficient_hydrogen})  and (\ref{eq:Stark_mixing_coefficient_hydrogen})--(\ref{eq:Stark_mixing_coefficient_hydrogen_reduced}) to obtain $\eta_{w}$ and $\eta_{S}$.

By applying the many--electron wavefunction (\ref{eq:wave_functions_2s_li_like}) one can derive the matrix element for the ${\rm 1s}^2 \; {\rm 2s_{1/2}} \to {\rm 1s}^2 \; {\rm 3s_{1/2}}$ transition:
\begin{eqnarray}
    \label{eq:transition_matrix_element_2s-3s_final}
    {\cal M}_{M_i M_f}({\rm 2S - 3S}) && \nonumber \\[0.2cm]
    && \hspace*{-2.5cm} = \frac{i}{4 \pi} \left(a_{M1} + 
    \lambda \left\{\eta^{(3) *}_w a^{(\rm{2s-3p})}_{E1} - \eta^{(2)}_w a^{(\rm{2p-3s})}_{E1} \right\} \right) \,
    \mathcal{D}^{0}_{M_i M_f \lambda} \nonumber \\[0.2cm]
    && \hspace*{-2.5cm} - \lambda \, \mathcal{E} \left( \tilde{\eta}^{(3) *}_S \, a^{(\rm{2s-3p})}_{E1} \, 
    \mathcal{D}^{1}_{M_i M_f \lambda} + \tilde{\eta}^{(2)}_S \, a^{(\rm{2p-3s})}_{E1} \, 
    \mathcal{D}^{1*}_{M_f M_i \, -\lambda} \right) \, ,
\end{eqnarray}
where the function $\mathcal{D}^{\nu}_{M_i M_f \lambda}$ is given by Eq.~(\ref{eq:D_function}).
This matrix element has a similar structure to its one--electron analog (\ref{eq:transition_matrix_element_1s-2s_final}) but depends on four coefficients $\eta^{(2)}_w \equiv \eta_w(\rm{2s-2p})$, $\eta^{(3)}_w \equiv \eta_w(\rm{3s-3p})$, $\tilde{\eta}^{(2)}_S \equiv \tilde{\eta}_S(\rm{2s-2p})$, and $\tilde{\eta}^{(3)}_S \equiv \tilde{\eta}_S(\rm{3s-3p})$ that describe weak--interaction and Stark mixing of the ground and excited states as well as on three reduced matrix elements $a_{M1}$, $a^{\rm{(2s-3p)}}_{E1}$, and $a^{\rm{(2p-3s)}}_{E1}$ of the magnetic and electric dipole transitions.  

We are ready now to employ the matrix element (\ref{eq:transition_matrix_element_2s-3s_final}) for the evaluation of the circular dichroism parameter for the ${\rm 1s}^2 \; {\rm 2s}_{1/2} \to {\rm 1s}^2 \; {\rm 3s}_{1/2}$ transition. As discussed already above, one needs to specify a particular experimental setup for this analysis. For example, the circular dichroism (\ref{eq:circular_dicroism_definition}) for the transitions between Zeeman sublevels $\ketm{{\rm 1s}^2 \; {\rm 2s}_{1/2} \, M_i} \to \ketm{{\rm 1s}^2 \; {\rm 3s}_{1/2} \, M_f}$, and for the case of the ``orthogonal'' geometry (${\mathcal {\bm B}} \perp {\mathcal {\bm E}} \perp {\bm k}$), can be written as:
\begin{eqnarray}
    \label{eq:dichroism_2S_3S_sublevels}
    \mathcal{A}_{M_i M_f} &\approx& 2 \left( {\rm Re}\left(\eta^{(3)}_w\right) \frac{a^{(\rm{2s-3p)}}_{E1}}{a_{M1}} -  {\rm Re}\left(\eta^{(2)}_w\right) \frac{a^{(\rm{2p-3s})}_{E1}}{a_{M1}}\right) \nonumber \\
    &\times& {\mathcal F}({\mathcal E}) \, ,
\end{eqnarray}
and where the function
\begin{eqnarray}
   {\mathcal F}({\mathcal E}) && \nonumber \\
   && \hspace*{-1cm} = \Bigg( 1 
    - \frac{\mathcal{E}}{\sqrt{\pi}}\left[M_i {\rm Re}\left(\tilde\eta^{(2)}_s\right)\frac{a^{(\rm{2p-3s})}_{E1}}{a_{M1}} + M_f {\rm Re}\left(\tilde\eta^{(3)}_s\right) \frac{a^{(\rm{2s-3p})}_{E1}}{a_{M1}}\right] \Bigg) \nonumber\\[0.2cm] 
    && \hspace*{-1cm} \times  \, \Bigg(1 + \frac{{\mathcal E^2}}{4\pi}\left[ \left|{\tilde \eta}^{(3)}_s\right|^2 \left(\frac{a^{(\rm{2s-3p})}_{E1}}{a_{M1}}\right)^2 +  \left|{\tilde \eta^{(2)}}_s \right|^2 \left(\frac{a^{(\rm{2p-3s})}_{E1}}{a_{M1}}\right)^2 \right]\nonumber\\[0.2cm] 
    && \hspace*{-1cm} -  \frac{2\mathcal{E}}{\sqrt{\pi}}\left[M_i{\rm Re}\left(\tilde\eta^{(2)}_s\right)\frac{a^{(\rm{2p-3s})}_{E1}}{a_{M1}} +M_f{\rm Re}\left(\tilde\eta^{(3)}_s\right)\frac{a^{(\rm{2s-3p})}_{E1}}{a_{M1}} \right] \nonumber\\[0.2cm] 
    && \hspace*{-1cm} + \frac{2\mathcal{E}^2}{\pi} M_i M_f {\rm Re}\left(\tilde\eta^{(2)}_s\right){\rm Re}\left(\tilde\eta^{(3)}_s\right)\frac{a^{(\rm{2s-3p})}_{E1} a^{(\rm{2p-3s})}_{E1}}{a_{M1}^2} \Bigg)^{-1} \, ,
\end{eqnarray}
becomes unity for ${\mathcal E} = 0$.

\begin{figure}[t]
    \centering
    \includegraphics[width=0.99\linewidth]{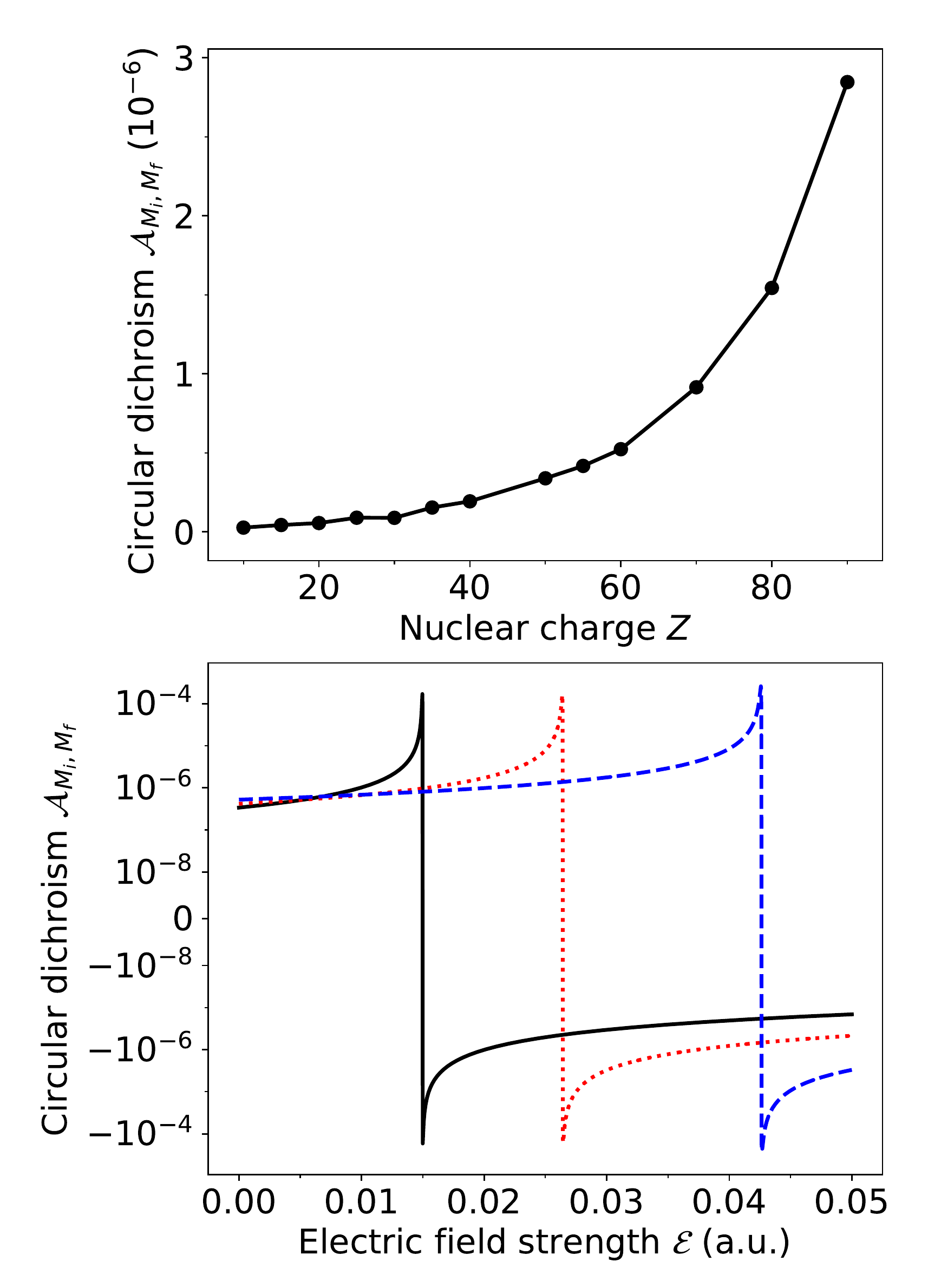}
    \caption{The circular dichroism (\ref{eq:dichroism_2S_3S_sublevels}) for the $\ketm{{\rm 1s}^2 \; {\rm 2s}_{1/2}\, M_i=+1/2} \to \ketm{{\rm 1s}^2 \; {\rm 3s}_{1/2}\, M_f=-1/2}$
    transition in lithium--like ions. In the upper panel, the parameter $\mathcal{A}_{M_i M_f}$ is displayed as a function of the nuclear charge $Z$ and for \textit{zero} electric field, $\mathcal{E} = 0$.  In the lower panel, in contrast, we display the variation of the dichroism with $\mathcal{E}$ for $Z$~=~50 (black solid line), 55 (red dashed line), and 60 (blue dash-dotted line). }
    \label{Fig4}
    \vspace{-0.5cm}
\end{figure}

Even though $\mathcal{A}_{M_i M_f}$ looks rather complicated, it has a similar structure to that of Eq.~(\ref{eq:dichroism_1s_2s_sublevels}). Indeed, it can be written as a product of (i) the ``weak--interaction'' part, given by the first line of Eq.~(\ref{eq:dichroism_2S_3S_sublevels}), and (ii) the function ${\mathcal F}({\mathcal E})$, which depends on the external electric field $\mathcal{E}$ and, hence, on the Stark mixing. One can note that the first (weak--interaction) part is just the circular dichroism for the case of vanishing field strength, $\mathcal{A}_{M_i M_f}({\mathcal E} = 0)$. We display it in the upper panel of Fig.~\ref{Fig4} as a function of $Z$. As seen from the figure, the circular dichroism $\mathcal{A}_{M_i M_f}({\mathcal E} = 0)$ steadily increases with the growth of $Z$ and lies in the range $\sim 10^{-7} - 10^{-6}$ for the medium--$Z$ to high--$Z$ domain. One can conclude that the effect of the weak--interaction mixing in lithium--like ions is comparable to that in the hydrogen--like ones. Since, moreover, the energy of the ${\rm 1s}^2 \; {\rm 2s}_{1/2} \to {\rm 1s}^2 \; {\rm 3s}_{1/2}$ transition does not exceed 300~eV even for heaviest ions \cite{BuC20}, and can be easily achieved at the Gamma Factory, the medium--$Z$ lithium--like ions can be considered as very promising candidates for parity--violation studies.

The circular dichroism (\ref{eq:dichroism_2S_3S_sublevels}) for transitions $\ketm{{\rm 1s}^2 \; {\rm 2s}_{1/2} \, M_i}$ $\to  \ketm{{\rm 1s}^2 \; {\rm 3s}_{1/2} \, M_f}$ also appears to be very sensitive to the variation of the strength of external electric field. As seen from the lower panel of Fig.~\ref{Fig4}, the parameter $\mathcal{A}_{M_i M_f}({\mathcal E})$, displayed as a function of ${\mathcal E}$, exhibits the resonant behaviour, similar to that was observed in Section~\ref{subsec:hlike_ions}. The amplitude of these ``resonances'', however, is much higher for the lithium--like ions. For example, the circular dichroism (\ref{eq:dichroism_2S_3S_sublevels}) for the excitation of $Z=50$ ion changes from $\mathcal{A}_{M_i M_f} = 3.4\times 10^{-7}$ for ${\mathcal E} = 0$ to $\mathcal{A}_{M_i M_f} = 6.5\times 10^{-5}$ for ${\mathcal E} = 1.49 \times 10^{-2}$ (a.u.). Again, such a high $\mathcal E$--sensitivity, caused by the weak--interaction---Stark interference, can be employed in Gamma Factory experiments for the accurate determination of $\eta^{(2)}_w$ and $\eta^{(3)}_w$ parameters.    

\begin{figure}
    \centering
    \includegraphics[width=0.99\linewidth]{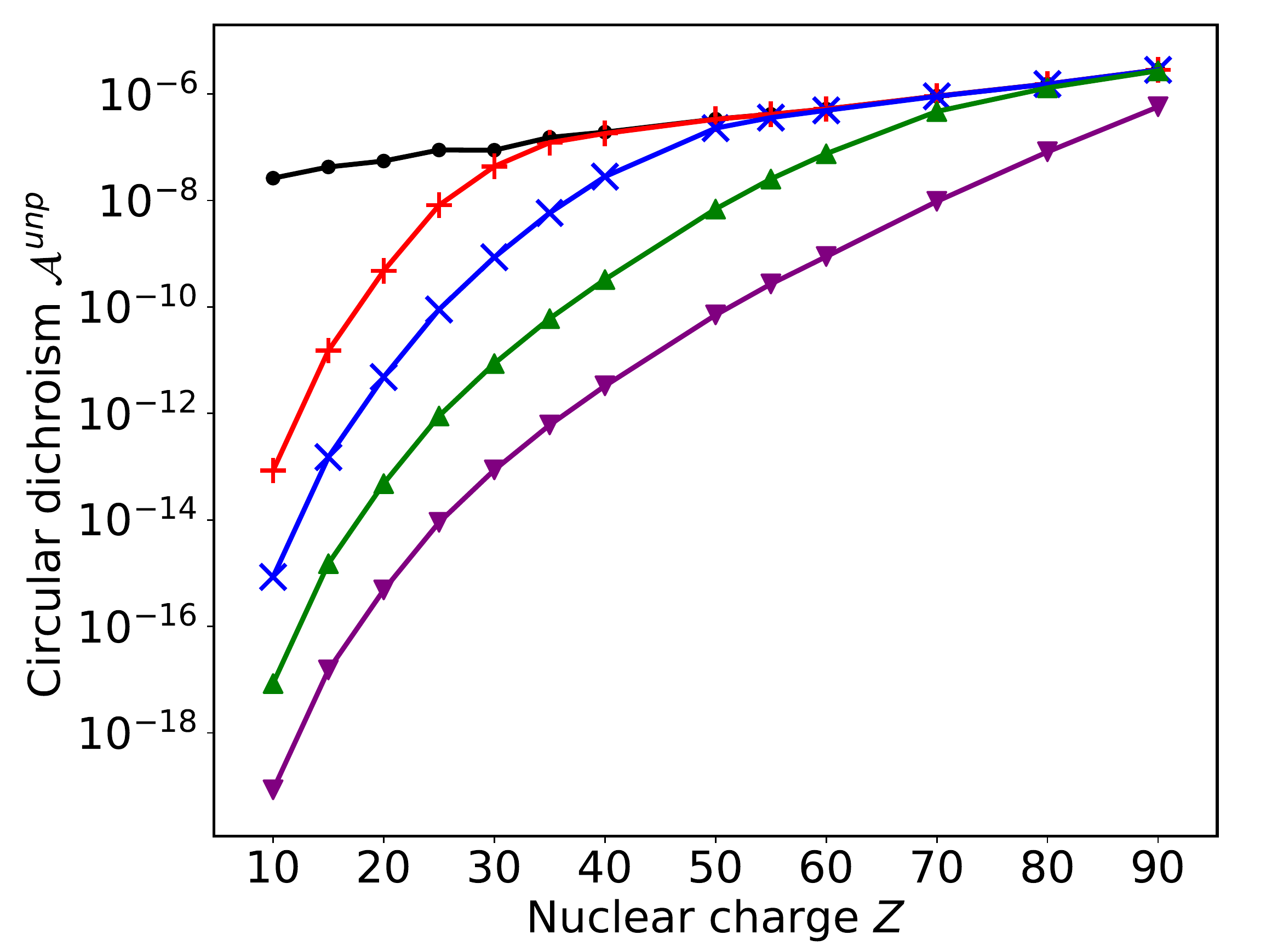}
    \caption{The circular dichroism parameters $\mathcal{A}^{\rm (unp)}$ for the ${\rm 1s}^2 \; {\rm 2s}_{1/2} \to {\rm 1s}^2 \; {\rm 3s}_{1/2}$ transition in lithium--like ions. The parameter $\mathcal{A}^{\rm (unp)}$ is given by Eq.~(\ref{eq:dichroism_2S_3S_unpolarized}) and is obtained for the case when magnetic sublevels of initial and final ionic states are unresolved. The calculations were performed for the external electric field with strength $\mathcal{E}$~=~0~a.u (black "$\bullet$"-line), 0.001~a.u. (red-"$+$" line), 0.01~a.u. (blue "x"-line), 0.1~a.u. (green "$\blacktriangle$"-line), and 1~a.u. (purple "$\blacktriangledown$"-line).}
    \label{Fig5}
\end{figure}

Similar to the discussion in Section \ref{subsec:hlike_ions}, we conclude our analysis of the parity--violation effects in lithium--like ions by considering ${\rm 1s}^2 \; {\rm 2s}_{1/2} \to {\rm 1s}^2 \; {\rm 3s}_{1/2}$ photoexcitation with \textit{unresolved} magnetic sublevel structure. The circular dichroism parameter can be derived for this case as: 
\begin{eqnarray}
    \label{eq:dichroism_2S_3S_unpolarized}
    \mathcal{A}^{\rm (unp)} &\approx& 2 \left( {\rm Re}\left(\eta^{(3) *}_w\right) \frac{a^{(\rm{2s-3p})}_{E1}}{a_{M1}} -  {\rm Re}\left(\eta^{(2)}_w\right) \frac{a^{\rm{(2p-3s)}}_{E1}}{a_{M1}}\right) \nonumber \\
    &\times& \frac{1}{1 + \frac{\left|\tilde{\eta}^{(2)}_s\right|^2 \mathcal{E}^2}{4\pi} \, \left(\frac{a^{(\rm{2p-3s})}_{E1}}{a_{M1}}\right)^2 + \frac{\left|\tilde{\eta}^{(3)}_s\right|^2 \mathcal{E}^2}{4\pi} \, \left(\frac{a^{(\rm{2s-3p})}_{E1}}{a_{M1}}\right)^2} \, .
\end{eqnarray}
As seen from this expression and Fig.~\ref{Fig5}, the absolute value of $\mathcal{A}^{\rm (unp)}$ is again maximal if no external electric field is applied and monotonically decreases with the rise of $\mathcal E$. For medium--$Z$ ions this ${\mathcal E}$--field--controlled reduction is rather large and can be used to extract the values of the weak--interaction mixing parameters.  

\section{Summary and outlook}
\label{sec:summary}

In summary, we have explored the feasibility of the Gamma Factory at CERN for the atomic parity violation studies with highly charged ions. A big advantage of the Gamma Factory setup is that it allows direct photoexcitation of medium-- and even high--$Z$ ions from their ground states. By measuring the rates of this excitation, if induced by right-- and left--circularly polarized light, one can determine the circular dichroism, caused by the \textit{mixing} of opposite--parity ionic levels. In turn, this mixing arises due to the weak interaction between electrons and nucleus, and can be further modified by applying external electric field.  

A theoretical analysis is reported investigating the weak--interaction and Stark mixing between s and p states and, hence, the circular dichroism of the 1s$_{1/2} \to$~2s$_{1/2}$ and ${\rm 1s}^2 \; {\rm 2s}_{1/2} \to {\rm 1s}^2 \; {\rm 3s}_{1/2}$ transitions in hydrogen-- and lithium--like ions, respectively. Based on relativistic calculations, we found that in the absence of the external electric field ${\mathcal E}$, the absolute value of the dichroism parameter is about $10^{-6}$ for both ionic species, and this value depends weakly on the nuclear charge $Z$. A further \textit{controllable} enhancement of the measured dichroism can be achieved by inducing transitions between magnetic Zeeman sublevels and for nonvanishing ${\mathcal E}$. We argue that external electric and magnetic fields of sufficient strength, demanded for this experimental scenario, can be produced at the Gamma Factory, which opens up a promising route for the future parity-violation studies with highly charged ions.

The present work is restricted to the most favorable geometry of the APV study, where the constant and homogeneous electric and magnetic fields are orthogonal to each other and to the direction of incident light. This simple case has to be reexamined for the realistic Gamma Factory experiments, where ultra--relativistic ions pass dipole bending as well as quadrupole and sextupole correction magnets. The analysis of such a realistic scenario depends on a particular experimental setup and will be presented elsewhere. \asc{The information about experimental details is also required to estimate the statistical sensitivity of the proposed APV experiments. This discussion is out of scope of the present---mainly theoretical---manuscript. We argue, moreover, that statistical sensitivity of APV studies at SPS and LHC facilities was carefully analyzed in Ref.~\cite{ZoB97}. The results, obtained in that work, when combined with the present findings will help to guide future experiments at the Gamma Factory.}

Further APV studies with highly charged ions present wide opportunities \cite{BuC20}: for instance, neutron radius of a nucleus may be evaluated from spin-independent APV mixing, and spin-dependent APV effects may be observed. Let us briefly discuss the theoretical background of both options. 

In Eq.\,\eqref{eq:PV_Hamiltonian} we assume neutron $\rho_n$ and proton $\rho_p$ nuclear densities to be equal ($\rho_n=\rho_p=\rho_N$), and neglect the neutron-skin contribution to $\hat{H}_w$. However, matrix elements of spin-independent PV Hamiltonian are, in fact, sensitive to the effective weak charge $\tilde{Q}_w$ which takes into account each nuclear density separately \cite{pollock_atomic_1992}:
\begin{equation}
    \tilde{Q}_w\approx -Nq_n+Zq_p(1-\sin^2\theta_W)\ .
\end{equation}
Here $q_n$ and $q_p$ are integrals of $\rho_n$ and $\rho_p$ together with the variation of electron wavefunction density $f({\bm r})$ at the nucleus:
\begin{equation}
    q_{n,p}=\int\rho_{n,p}({\bm r})f({\bm r})d^3r\ .
\end{equation}
The difference in root-mean-square radii between neutron and proton nuclear densities is referred to as \textit{neutron skin}. Since $1-\sin^2\theta_W\approx 0.08$, effective weak charge is primarily determined by the neutron distribution $\rho_n$. However, $\rho_n$ is poorly known in comparison to the proton density $\rho_p$: the latter can be measured, for example, in electron-scattering experiments \cite{angeli_table_2013}. In fact, APV is one of the few phenomena which can probe neutron distributions in nuclei. Hence measurements of PV effects in partially stripped ions could provide an invaluable source of information on neutron skins, especially given the simplicity of atomic structure calculations \cite{viatkina_dependence_2019}.

Additionally, several spin-dependent APV phenomena can be observed in atoms and ions with nuclear spin ${\bm I}\neq 0$. One of these effects, significant in heavier nuclei, is caused by nuclear anapole moment (NAM), which itself arises from weak interaction within the nucleus. The effect can be expressed through an effective Hamiltonian \cite{SaB18}:
\begin{equation}
    \hat{H}_\mathrm{NAP}({\bm r})=\frac{G_F}{\sqrt{2}}\eta_\mathrm{NAM}\left({\bm \alpha}\cdot{\bm I}\right)\rho_N({\bm r})\ ,
\end{equation}
where the operator ${\bm \alpha}$ ($\alpha_i=\gamma_0\gamma^i$) acts on electrons, and $\eta_\mathrm{NAM}$ signifies the strength of the nuclear-anapole effect. It scales as $\eta_\mathrm{NAM} \propto A^{2/3}$ with the atomic number $A$. The anapole moment for a nuclear current density ${\bm j(r)}$ is
\begin{equation}
    {\bm a}=-\pi\int r^2 {\bm j(r)}d^3{\bm r}=\frac{G_F}{\sqrt{2}|e|}\eta_\mathrm{NAM}{\bm I}\ .
\end{equation}
Nuclear anapole moment causes PV mixing, which could be observed, for instance, in helium-like ions in the transition between the $1s^2$~$^1S_0$,~$F=I$ and $1s2s$~$^1S_0$,~ $F=I$ levels \cite{BuC20}.

\asc{We would like to stress that PV measurements in highly charged ions offer a unique opportunity to probe a number of nuclear effects, e.g. those discussed above. The primary advantage of highly charged ions lies in their relatively simple electronic structure, where the relevant atomic properties can be calculated with much higher precision than is currently achievable in many-electron systems. Neutron-skin and anapole-moment effects are more pronounced in heavier atoms; however, precision atomic calculations necessary for interpreting  measurements in heavy many-electron atoms may pose an insurmountable problem.
Indeed, this problem is the primary motivation behind isotopic chain measurements, where the ratio of APV effects in two isotopes of the same atom is taken in order to cancel out the atomic factors \cite{dzuba_enhancement_1986, antypas_isotopic_2019}. However, in doing so, one amplifies
nuclear uncertainties \cite{fortson_nuclear-structure_1990}. A single-isotope PV measurement in a system with simple electronic structure is a great asset for determining the nuclear effects in question. Moreover, the LHC-based setup of Gamma Factory potentially allows for handling highly charged ions with unstable nuclei, which may exhibit enhanced PV effects e.g. due to a pronounced nuclear deformation \cite{flambaum_enhancing_2016,flambaum_effect_2017}}.

In short, APV experiments at Gamma Factory are a promising way to probe the properties of medium and heavy nuclei, as well as fundamental characteristics of weak interaction. More specifically, these experiments may provide a unique opportunity to measure both nuclear-spin-independent and spin-dependent PV effects in highly charged ions.

\section*{Acknowledgements}

The authors are grateful for helpful discussions with Mieczyslaw Witold Krasny and Mikhail Kozlov. This work was funded by the Russian Science Foundation (Grant No. 19-12-00157), the GFK-Fellowship (JGU Mainz), and under Germany’s Excellence Strategy—EXC-2123 QuantumFrontiers—390837967. The work at Mainz was supported by the Deutsche Forschungsgemeinschaft (DFG) - Project ID 423116110 and by the Cluster of Excellence Precision Physics, Fundamental Interactions, and Structure of Matter (PRISMA+ EXC 2118/1) funded by the DFG within the German Excellence Strategy (Project ID 39083149).

%
%
\bibliographystyle{andp2012}
\bibliography{b20.parity_violation_GF}

\end{document}